\documentstyle[emulateapj,danonecolfloat]{article}

\newcommand{\kms}{\ifmmode{{\rm ~km~s}^{-1}}\else{~km~s$^{-1}$}\fi}

\newcommand{\mum}{\mu \rm m}

\newbox\grsign \setbox\grsign=\hbox{$>$} \newdimen\grdimen 
\grdimen=\ht\grsign
\newbox\laxbox \newbox\gaxbox
\setbox\gaxbox=\hbox{\raise.5ex\hbox{$>$}\llap
     {\lower.5ex\hbox{$\sim$}}}\ht1=\grdimen\dp1=0pt
\setbox\laxbox=\hbox{\raise.5ex\hbox{$<$}\llap
     {\lower.5ex\hbox{$\sim$}}}\ht2=\grdimen\dp2=0pt

\slugcomment{Accepted for Publication in {\it The Astrophysical Journal} (2003)}

\begin{document}

\twocolumn[
\title{Radio and Far-Infrared Emission as Tracers of Star Formation and AGN 
in Nearby Cluster Galaxies}

\author{Naveen A. Reddy}
\affil{California Institute of Technology, Astronomy Department MS 105-24, 
Pasadena, CA 91125; nar@astro.caltech.edu}
\and
\author{Min S. Yun}
\affil{University of Massachusetts, Department of Astronomy, 522 LGRT,
Amherst, MA 01003; myun@astro.umass.edu}

\begin{abstract}

We have studied the radio and far-infrared (FIR) emission from $114$ galaxies in the $7$ nearest 
clusters ($<100$ Mpc) with prominent X-ray emission to investigate the impact 
of the cluster environment on the star formation and AGN activity in the member galaxies.  
The X-ray selection criterion is adopted to focus on the most massive and dynamically relaxed clusters.
A large majority of cluster galaxies show an excess in radio emission over that predicted from 
the radio-FIR correlation, the fraction of sources with radio excess increases toward cluster cores, 
and the radial gradient in the FIR/radio flux ratio is a result of radio enhancement.  
Of the radio-excess sources, $70\%$ are early-type galaxies and the same fraction host an AGN.  
The galaxy density drops by a factor of 10 from the composite cluster center out to $1.5$~Mpc,
yet galaxies show no change in FIR properties over this region, and show no indication of 
mass segregation.  
We have examined in detail the physical mechanisms that might impact the FIR and radio emission 
of cluster galaxies.  While collisional heating of dust may be important for galaxies in cluster
centers, it appears to have a negligible effect on the observed FIR emission for our sample galaxies.
The correlations between radio and FIR luminosity and radius 
could be explained by magnetic compression from thermal ICM pressure.  We also find that simple 
delayed harassment cannot fully account for the observed radio, FIR, and mid-IR properties of 
cluster galaxies.

\end{abstract}  

\keywords{galaxies: clusters: general --- galaxies: clusters: individual (A262, AWM7, A426, A1060, A1367, Virgo, 
A1656) --- galaxies: elliptical and lenticular, cD --- galaxies: interactions --- infrared: galaxies --- radio continuum: galaxies}]

\section{Introduction}

The radio and far-infrared (FIR) properties of local galaxies with a wide range in bolometric 
luminosity, from the faintest dwarf galaxies to ultraluminous infrared galaxies (ULIRGs), 
have been well documented.  The overwhelming majority of these galaxies obey the linear 
correlation between radio and FIR luminosity which holds over 5 orders of magnitude in 
luminosity (Price \& Duric 1992).  The mechanisms for radio and FIR emission are thought 
to be well understood and both have been shown to be accurate tracers of recent 
star formation (see Yun, Reddy, \& Condon 2001--hereafter YRC--and references
therein).  The small percentage of luminous galaxies which deviate from the FIR-radio 
relation in the local universe are thought to be powered by AGN (for those showing excess radio emission over 
that expected from the radio-FIR correlation and excess mid-IR emission) or are in 
a short-lived dust enshrouded AGN/starburst phase (for those showing excess FIR emission; YRC).  
Deviations from the correlation may also arise from environmental changes such as the presence 
of an intracluster medium (ICM).  Studies have found notably lower FIR-radio 
flux ratios for cluster galaxies than expected from the FIR-radio relation in the field 
(see Miller \& Owen 2001a), indicating some modification of the radio and/or FIR emission mechanisms 
in cluster galaxies.  

Mechanisms that can only affect the observed FIR emission include collisional heating of dust 
grains by ICM electrons and ram pressure stripping of ISM.  In the latter case, observations indicate 
that such stripping only strongly affects the outer HI disk of the galaxy, leaving the dense 
molecular gas and dust component intact (Kenney \& Young 1986, Cayette et al. 1994).  A consequence 
of the lower star formation rates observed in cluster galaxies (e.g, Gomez et al. 2002) is a 
corresponding reduction in both FIR and radio emission.  Enhancement of radio continuum and 
FIR emission may result from tidally induced star formation.  Mechanisms affecting the 
observed radio emission include an increase in the magnetic field strength via gas compression 
as galaxies move through the ICM (Gavazzi \& Boselli 1999; Rengarajan, Karnik, \& Iyengar 1997; 
Anderson \& Owen 1995; Scodeggio \& Gavazzi 1993), simple thermal compression of the magnetic 
field (e.g., Miller \& Owen 2001a), and/or elevated AGN activity arising from frequent 
galaxy-galaxy collisions or galaxy harassment in the cluster environment (e.g., Moss \& Whittle 1993; 
Lake, Katz, \& Moore 1998).  Studying radio and FIR emission should therefore be useful in examining
these mechanisms.

Miller \& Owen (2001a) conducted a study of the radio and FIR properties of 
a sample of 329 local Abell cluster galaxies.  Optically-identified AGNs in their
sample show greater radio-to-FIR flux ratios with more scatter than star forming
galaxies and galaxies in cluster cores show enhanced radio emission.  While unable to 
rule out ram pressure enhancement, they find that thermal compression of the magnetic field
alone could produce the observed radio luminosity segregation.

Prior investigations into the effects of the ICM on galaxy properties have primarily selected
clusters based on their optical properties that may not accurately 
reflect the state of the intracluster medium.  For example, the Miller \& Owen (2001b) cluster
sample has 329 spectroscopically confirmed cluster galaxies, half of which reside in poor
clusters where the environmental effects are not as pronounced as in rich clusters. 
We have adopted a different approach in selecting 
clusters based on the presence of X-ray emission.  The most common explanation for this emission in 
clusters, first proposed by Felten et al. (1966), is that X-rays result from thermal bremsstrahlung 
from hot ($\approx 10^{8}$ K), low density ($\sim 10^{-3}$ atoms~cm$^{-3}$) intracluster gas.  
The inclusion of X-ray luminous clusters in our sample (1) ensures that only massive, 
dynamically-relaxed clusters are chosen, and (2) is immune to the biases inherent in optically-selected 
surveys, such as sensitivity to morphological type, AGN fraction, and ages and masses of galaxies.

X-ray cluster and radio/FIR galaxy selection allows us to compile a homogeneous sample 
of cluster galaxies in order to address the key questions of this study: (1) does the 
rich cluster environment help or hinder star formation activity in cluster galaxies, and 
(2) how does the presence of an ICM affect the frequency of AGN sources?  To aid in 
addressing these questions, we also examine the following: convolution due to projection 
effects (\S 3), clustering of early-type sources as predicted by the density-morphology relation
(Dressler 1980; \S 6.1), and FIR emissivity as a function of mass (\S 6.2). 
All of these effects can influence the observed radio and FIR properties of cluster galaxies and must
be carefully disentangled.  To determine whether rich clusters preferentially affect low or
high luminosity galaxies, we also examine the faint and bright end behavior of the cluster
luminosity function (LF) and compare with that found in the field.

The organization of this paper is as follows.  In \S 2, we present the criteria used
to select clusters and individual galaxies.  The effects of projection are quantified
in \S 3.  The radio-FIR relation and flux density ratio for cluster galaxies are
discussed in \S 4.  The effects of the ICM on the galaxy radio and IR luminosity functions
are discussed in \S 5.  The radio and FIR properties of
the sample are interpreted in the context of the density-morphology relation, mass segregation,
and various emission mechanisms in \S 6, including electron-dust collisional heating and ram-pressure
and thermal magnetospheric compression.  We also examine the mid-IR properties of the cluster
sample to address the frequency of AGN sources in \S 6.

\section{Sample Selection}

\subsection{Cluster Selection}

The clusters used in this study were culled from the flux-limited sample of $55$ bright X-ray
clusters compiled by Edge et al. (1990) using the following selection criteria.  First, 
we chose only those clusters with $\delta \geq -40^{\circ}$, corresponding 
to the survey limit of the NRAO VLA Sky Survey (NVSS) from which radio data were
obtained.  Clusters near the galactic plane ($|b|< 10^{\circ}$) were also excluded.  
Finally, a redshift cutoff of $z<0.025$ was applied ($100$ Mpc for H$_{\rm o}=75$ 
km~s$^{-1}$~Mpc$^{-1}$---this value of H$_{\rm o}$ is assumed throughout the paper), to ensure 
accurate NVSS and IRAS detections given the sensitivity and confusion limits of the two surveys.  
As an added advantage, we only select nearby, well-studied clusters, so that a case-by-case 
study of individual galaxies can be done.  This selection process yielded $7$ clusters: 
A262, AWM7, A426 (Perseus), A1060 (Hydra), A1367, Virgo, and A1656 (Coma).  The ACO (Abell, Corwin,
\& Olowin 1989) cluster
richness among these $7$ clusters ranges from -1 to 2, but all have virial masses 
of $\sim 10^{14}$ M$_{\odot}$ to $\sim 10^{15}$ M$_{\odot}$ (David, Jones, \& Forman 1996; Koranyi \& Geller 2000; Ettori, 
Fabian, \& White 1998; Tamura et al. 2000; Donnelly et al. 1998; Colless \& Dunn 1996; Schindler, 
Binggeli, \& B\"{o}hringer 1999) indicative of a well-defined potential.  The central position, 
assumed redshift, and $2-10$ keV X-ray luminosity of each cluster are included in the data 
summarized in Table~\ref{tab:clsum}.

\begin{deluxetable}{lcccccccccc}
\tablewidth{0pc}
\tablecaption{Cluster Properties}
\tablehead{
\colhead{}                  &
\colhead{$\alpha$}                &
\colhead{$\delta$}                &
\colhead{$<cz>$\tablenotemark{a}}                   &
\colhead{$\sigma$\tablenotemark{b}}        &
\colhead{}                 &
\colhead{N$_{Core}$}       &
\colhead{N$_{Ring}$}      &
\colhead{}        &
\colhead{}         &
\colhead{$2-10$ keV L$_x$\tablenotemark{d}}                 \\
\colhead{Name}                         &
\colhead{(J2000.0)}           &
\colhead{(J2000.0)}           &
\colhead{(km~s$^{-1}$)}                   &
\colhead{(km~s$^{-1}$)}                   &
\colhead{Class\tablenotemark{c}}                              &
\colhead{(Obs)}               &
\colhead{(Obs)}               &
\colhead{N$_{early}$}                &
\colhead{N$_{late}$}                     &
\colhead{($10^{43}$h$^{-2}_{0.75}$ erg~s$^{-1}$)}}
\startdata
A262 & 01 52 50.4 & 36 08 46 & 4887 & 588 & 0 & 6 & 7 & 5  & 8  & $1.22$ \\
AWM7 & 02 54 32.2 & 41 35 10 & 5168 & 680 & -1 & 1 & 4 & 2 & 3 & $5.20$ \\
A426 & 03 18 36.4 & 41 30 54 & 5366 & 1324 & 2 & 1 & 5 & 1  & 5  & $48.9$ \\
A1060 & 10 36 51.3 & -27 31 35 & 3777 & 647 & 1 & 5 & 16 & 4  & 17  & $1.09$ \\
A1367 & 11 44 29.5 & 19 50 21 & 6595 & 879 & 2 & 11 & 9 & 8  & 12  & $3.08$ \\
Virgo & 12 26 32.1 & 12 43 24 & 1075 & 632 & 1 & 3 & 20 & 2  & 21  & 0.52\\
Coma & 12 59 48.7 & 27 58 50 & 6925 & 1008 & 2 & 6 & 20 & 11 & 15  & $33.3$ \\
\enddata
\tablenotetext{a}{Systemic cluster velocity from NED}
\tablenotetext{b}{Velocity dispersion ($1\sigma$) from Strubble \& Rood (1999),
Koranyi et al. (1998), and Fadda et al. (1996)}
\tablenotetext{c}{Richness class from Abell, Corwin, \& Olowin (1989)}
\tablenotetext{d}{$2-10$ kev luminosity from Edge et al. (1990)} 
\label{tab:clsum}
\end{deluxetable}

\subsection{Galaxy Selection}

Individual galaxies in the $7$ aforementioned clusters were selected using the following
procedure.  We used the NASA/IPAC Extragalactic Database (NED) to search for all cataloged
galaxies with available redshift within $1.5$ Mpc projected distance from the 
cluster center.  Sources with radial velocities $>2000$ km~s$^{-1}$ than
the systemic cluster redshift were excluded from the sample.  
NED selection is subject to the availability of data, and to investigate this bias
we applied the selection criteria above to the Miller \& Owen (2001b) sample
for clusters common between their sample and ours: A262, A426, 
A1367, and Coma.  The same was done with the AWM7, A1060, and Virgo 
homogenous samples of Koranyi \& Geller (2000), Richter (1989), and Binggeli, 
Sandage, \& Tammann (1985), respectively.  All of these samples reduce to subsets of
our sample after imposing our selection criteria.  There are $3$, $2$, and $5$
galaxies in A262, A1367, and Coma, respectively, which are not included in the
homogeneous samples above, and may be drawn from other surveys having
potentially different selection biases.  However, we note that these sources
have radio and infrared (IR) luminosities comparable to the rest of the sample, and are
morphologically diverse (half are early- and half are late-type galaxies), so their 
inclusion in the sample should not significantly affect our conclusions.

A constant velocity cutoff of $2000$ km~s$^{-1}$ preferentially
excludes sources with high peculiar velocity, particularly for A1060 and Coma,
both having line-of-sight velocity dispersions of $\sigma \sim 1000$ km~s$^{-1}$.  Adopting
a $3\sigma$ cutoff velocity for A1060 and Coma adds only $1$ source with
detected radio and FIR flux to the sample, the starburst galaxy NGC4858.
Our velocity cutoff quantitatively only affects the normalization of the cluster
luminosity function (\S 5), since there is evidence that all types of galaxies
fill luminosity space uniformly (i.e., no mass segregation; \S 6.2).

The FIR and radio data were obtained from the Infrared Astronomical Satellite 
(IRAS; Neugebauer et al. 1984) and the NRAO VLA Sky survey (NVSS; Condon et al. 1998), 
respectively.  The IRAS data include $12\mu$m, $25\mu$m, $60\mu$m, and $100\mu$m flux 
measurements with a 1$\sigma$ error ellipse of $15''$.  The total luminosity between $40\mu$m 
and $120\mu$m, dominated by young massive stars, can be estimated from the 
$60\mu$m and $100\mu$m flux.  The NVSS radio survey at $1.4$ GHz 
($20$ cm) with the D and DnC configurations of the Very Large Array, has an angular 
resolution of $45''$ with a $1\sigma$ sensitivity of $0.5$ mJy~beam$^{-1}$ at $20$ cm.  
The NVSS is sensitive to galaxies with star formation rates $\ga 1.4$ M$_{\odot}$~yr$^{-1}$ 
within a distance of $100$ Mpc (YRC; Condon 1992).

In the case of the two nearest clusters, A1060 and Virgo, we searched the NVSS database for radio
sources within $10$ kpc projected distance (the typical size of a disk galaxy)
from the optical position of each of the galaxies.  For the other $5$ clusters, 
we chose a search radius of $30''$ based upon the NVSS and IRAS error ellipses.  
The probability of finding an unrelated NVSS source within the search radius is 
$<0.1\%$. In some cases, confusion between a foreground galaxy and a 
background radio source (e.g., UGC1416), or between interacting companions 
(NGC4567 and NGC4568), was resolved using a more detailed inspection.  
For nearby galaxies with large angular extent in Virgo, the NVSS underestimates the 
flux due to extended disk emission and in some cases produces
multiple radio matches for a single optical position.  In these cases, 
corrections to the radio flux were obtained from the multiple-snapshot 
study of Condon (1987).

IRAS SCANPI coadditions of each 
galaxy were obtained and the source is included in our sample if the IRAS source 
centroid is within $30''$ of the nominal radio position from the NVSS database and 
has $S/N > 3$ in the $60\mu$m and $100\mu$m bands.  The IRAS satellite 
has a raw resolution of $\sim 4'$ and for extended sources greater than this, the SCANPI 
software can miss extended flux.  There are $24$ such sources in Virgo.  
IRAS fluxes for $14$ of these extended sources were obtained from the IRAS Bright Galaxy 
imaging compilation of Soifer et al. (1989).  Accurate fluxes could not be found in the literature
for $10$ of the extended objects in Virgo, but all of these sources are excluded from the 
sample after adopting a luminosity cutoff (see \S 2.3).  To summarize, we selected galaxies
from NED with redshifts and they are included in the sample if they have secure NVSS and
IRAS $60\mum$ and $100\mum$ detections.  This selection process yields
$182$ galaxies in the composite cluster.

The $1.4$ GHz radio and $60\mu$m luminosities were calculated using the following relations:
$$\log L_{1.4 \rm GHz} ({\rm W~Hz}^{-1}) = 20.08 + 2\log D + \log S_{1.4\rm GHz}\eqno(1)$$
$$\log L_{60\mum} (L_\odot) = 6.014 + 2\log D + \log S_{60\mum}\eqno(2)$$
where D is the distance in Mpc, and  
$S_{1.4 \rm GHz}$ and $S_{60\mum}$ are flux densities in units of Jy.  
The $60\mu$m luminosity is related to the FIR luminosity, $L_{\rm FIR}$, by
$$L_{\rm FIR} (L_\odot)
=\left(1+{{S_{100\mum}}\over{2.58 S_{60\mum}}}\right) L_{60\mum}\eqno(3)$$
(see Helou et al. 1988).  The typical uncertainty in the luminosities is $\sim 15\%$
($\sigma_{\rm \log L}\sim 0.065$), based
on NVSS and IRAS absolute flux calibration errors.  Intrinsic scatter due to variations 
in FIR and radio emission dominates the uncertainties in the flux scaling (YRC).

\subsection{Sample Limits and Divisions}

\begin{deluxetable}{llcccccccl}
\tablewidth{0pc}
\tablecaption{Cluster Galaxies in the Luminosity-Limited Sample\label{tab:clgal}}
\tablehead{
\colhead{}            &
\colhead{}               &
\colhead{$\alpha$}           &
\colhead{$\delta$}           &
\colhead{m}                  &
\colhead{$<cz>$}             &
\colhead{$\log L_{1.4GHz}$}   &
\colhead{$\log L_{60\mum}$}  &
\colhead{}                  &
\colhead{}              \\
\colhead{Cluster}                   &
\colhead{Name}                   &
\colhead{(2000.0)}           &
\colhead{(2000.0)}           &
\colhead{(mag)}              &
\colhead{(km~s$^{-1}$)}      &
\colhead{(W~Hz$^{-1}$)}      &
\colhead{(L$_{\odot}$)}      &
\colhead{q}                   &
\colhead{Notes}                   }
\startdata
A262 &  NGC0688              &  1  50 44.41 &  35 17 15.1 &     13.35 &     4151  &  21.68 & 9.76 & 2.37 &  (R')SAB(rs)b;Sbrst   \\
 & UGC1308              &  1  50 51.34 &  36 16 34.3 &     13.77 &     5171  &  22.01 & 9.00 & 1.29 &  E                    \\
 & UGC1319              &  1  51 28.99 &  36  3 56.7 &     14.50  &     5310  &  21.95 & 9.63 & 2.03 &  Irr                  \\
 & NGC703               &  1  52 39.30 &  36 10 16.0 &     14.27 &     5580  &  21.63 & 9.09 & 1.91 &  S0-                  \\
& UGC1347              &  1  52 45.98 &  36 37 8.1 &     13.49 &     5543  &  21.79 & 9.82 & 2.36 &  SAB(rs)c             \\
& NGC708               &  1  52 46.45 &  36  9 8.3 &     13.70  &     4855  &  22.53 & 9.10 & 0.98 &  cD;E;Sy2             \\
& NGC710               &  1  52 54.25 &  36  3 12.1 &     14.27 &     6125  &  22.25 & 9.82 & 1.84 &  Scd                  \\
& KUG0151+356          &  1  54 11.72 &  35 54 57.0 &     15.70  &     4387  &  21.11 & 8.99 & 2.36 &  Spiral               \\
& UGC1385              &  1  54 53.81 &  36 55 4.3 &     13.90  &     5621  &  21.98 & 10.43 & 2.64 &  (R)SB0/a;Sbrst       \\
& UGC1416              &  1  56 46.18 &  36 53 3.3 &     14.72 &     5470  &  21.11 & 8.97 & 2.17 &  Sb                   \\
& NGC753               &  1  57 41.99 &  35 54 58.8 &     12.97 &     4903  &  22.23 & 10.20 & 2.31 &  SAB(rs)bc            \\
& NGC759               &  1  57 50.43 &  36 20 37.2 &     13.84 &     4667  &  21.89 & 9.60 & 2.00 &  E                    \\
& UGC1460              &  1  59 4.61 &  36 15 25.2 &     14.64 &     4874  &  21.24 & 9.50 & 2.53 &  Sa                   \\
\\
AWM7 &  IC259                &  2  49 45.80 &  41  3 7.7 &     15.00    &     6108  &  21.50 & 9.74 & 2.52 &  SB0?                 \\
& NGC1106              &  2  50 40.49 &  41 40 17.8 &     13.30  &     4337  &  22.88 & 9.78 & 1.10 &  SA0+;Sy2             \\
& UGC2350              &  2  52 40.27 &  41 23 45.8 &     14.60  &     3964  &  21.22 & 9.38 & 2.51 &  Sb                   \\
& NGC1122              &  2  52 50.76 &  42 12 18.1 &     12.90  &     3599  &  21.80 & 10.03 & 2.53 &  SABb                 \\
& CGCG539-121          &  2  54 16.65 &  42 43 34.1 &     14.70  &     6331  &  21.91 & 9.82 & 2.18 &  Spiral               \\
\\
A426 & UGC2608              &  3  15 1.48 &  42  2 8.8 &     13.70  &     6998  &  22.76 & 10.63 & 2.06 &  (R')SB(s)b;Sy2       \\
& UGC2617              &  3  16 1.32 &  40 53 14.7 &     13.80  &     4697  &  21.33 & 9.53 & 2.69 &  SAB(s)d              \\
& IC310                &  3  16 42.95 &  41 19 28.7 &     13.89 &     5660  &  23.01 & 9.50 & 0.86 &  SA(r)                \\
& UGC2654              &  3  18 43.38 &  42 17 59.9 &     14.20  &     5793  &  21.87 & 9.94 & 2.38 &  S?                   \\
& NGC1275              &  3  19 48.16 &  41 30 42.1 &     12.64 &     5264  &  25.15 & 10.52 & -0.47 &  cD;pec;NLRG;Sy2      \\
& UGC2696              &  3  22 5.79 &  42 10 15.3 &     15.50  &     5453  &  21.25 & 9.09 & 2.13 &  Scd                  \\
\\
A1060 &  IRAS10288-2824       &  10 31 13.55 &  -28 39 51.8 &     15.33 &     3690  &  20.98 & 9.20 & 2.43 &  SABa                 \\
& ESO436-G034          &  10 32 44.35 &  -28 36 40.2 &     14.38 &     3624  &  21.21 & 9.32 & 2.50 &  Sb                   \\
& ESO501-G017          &  10 34 24.09 &  -26 29 27.8 &     14.97 &     4429  &  21.17 & 9.46 & 2.47 &  S0(7)                \\
& NGC3285B             &  10 34 37.22 &  -27 39 12.5 &     13.86 &     2952  &  21.09 & 9.08 & 2.39 &  SB(rs)b              \\
& ESO436-IG042         &  10 34 38.74 &  -28 35 0.3 &     14.40  &     3190  &  21.77 & 9.87 & 2.28 &  E4                   \\
& ESO436-G046          &  10 34 50.35 &  -28 35 4.7 &     13.42 &     3436  &  20.92 & 9.60 & 2.97 &  SB(rs)b              \\
& ESO437-G002          &  10 34 59.31 &  -28  4 44.2 &     14.63 &     2311  &  21.30 & 9.23 & 2.28 &  Sc;pec               \\
& ESO437-IG003         &  10 35 7.89 &  -27 59 27.2 &     ...     &     2391  &  21.22 & 8.95 & 2.15 &  ...                    \\
& ESO437-G004          &  10 35 23.75 &  -28 18 49.8 &     13.93 &     3304  &  21.05 & 9.21 & 2.53 &  SB(r)b               \\
& ESO437-G015          &  10 36 57.88 &  -28 10 31.2 &     13.53 &     2743  &  20.99 & 9.09 & 2.32 &  SB(s);sp             \\
& NGC3312              &  10 37 2.55 &  -27 33 52.7 &     12.68 &     2886  &  22.17 & 9.30 & 1.56 &  Sab(r)p;AGN          \\
& ESO501-G045          &  10 37 12.64 &  -26 40 22.5 &     15.11 &     4578  &  20.92 & 9.17 & 2.52 &  S0(3)/a(r)           \\
& NGC3314A             &  10 37 12.64 &  -27 41 3.1 &     14.40  &     2872  &  21.91 & 9.60 & 1.99 &  Sab:sp               \\
& ESO501-G053          &  10 37 37.91 &  -26 16 37.0 &     14.80  &     3814  &  21.06 & 9.33 & 2.56 &  Sa                   \\
& ESO501-G059          &  10 37 49.46 &  -27  7 17.4 &     14.97 &     2437  &  21.66 & 9.36 & 1.99 &  Sc                   \\
& ESO501-G065          &  10 38 33.53 &  -27 44 12.4 &     13.71 &     4425  &  21.86 & 9.55 & 2.00 &  SB(s)d:pec           \\
& ESO437-G025          &  10 38 40.10 &  -28 34 4.9 &     14.49 &     3458  &  20.98 & 9.19 & 2.57 &  S(B)b                \\
& IRAS10369-2827       &  10 39 16.78 &  -28 42 52.1 &     15.69 &     2811  &  21.12 & 9.17 & 2.24 &  S0/a                 \\
& ESO501-G068          &  10 39 17.47 &  -26 50 23.7 &     14.25 &     3090  &  20.93 & 9.15 & 2.57 &  S(B)bc(rs)           \\
& NGC3336              &  10 40 17.15 &  -27 46 46.4 &     13.00    &     4000  &  21.72 & 9.59 & 2.22 &  SA(rs)c              \\
& ESO501-G075          &  10 40 58.75 &  -27  5 8.8 &     13.48 &     4750  &  21.18 & 9.33 & 2.56 &  SA(s)c               \\
\\
A1367 & CGCG097-068          &  11 42 24.49 &  20  7 11.8 &     14.55 &     5974  &  22.05 & 10.15 & 2.35 &  Sbc                  \\
& CGCG097-073          &  11 42 56.22 &  19 58 10.3 &     15.50  &     7275  &  22.30 & 9.37 & 1.29 &  SAcd:pec             \\
& UGC6680              &  11 43 2.27 &  19 38 36.1 &     15.04 &     6985  &  21.79 & 8.94 & 1.71 &  Sb                   \\
& CGCG097-079          &  11 43 12.22 &  20  0 26.8 &     15.70  &     7000  &  21.67 & 9.38 & 2.03 &  Irr                  \\
& UGC6697              &  11 43 48.45 &  19 58 14.3 &     14.08 &     6725  &  22.70 & 10.10 & 1.66 &  Im                   \\
& CGCG097-092          &  11 43 58.49 &  20 11 16.3 &     15.50  &     6373  &  21.43 & 9.52 & 2.33 &  Sbc                  \\
& NGC3840              &  11 43 59.28 &  20  4 32.5 &     14.54 &     7368  &  21.60 & 9.79 & 2.45 &  Sa                   \\
& NGC3842              &  11 44 1.81 &  19 56 56.7 &     12.78 &     6316  &  22.06 & 9.44 & 1.79 &  E                    \\
& IRAS11419+2022       &  11 44 32.05 &  20  6 22.0 &     15.40  &     7230  &  21.46 & 9.94 & 2.70 &  S0                   \\
& UGC6719              &  11 44 47.47 &  20  7 25.5 &     14.44 &     6571  &  21.42 & 9.30 & 2.15 &  Sab                  \\
& NGC3860B             &  11 44 46.80 &  19 46 13.3 &     15.30  &     8293  &  21.73 & 9.81 & 2.45 &  Irr                  \\
& NGC3860              &  11 44 48.86 &  19 47 47.1 &     14.22 &     5595  &  22.16 & 9.75 & 1.98 &  Sa                   \\
& NGC3859              &  11 44 51.20 &  19 27 21.5 &     14.76 &     5468  &  21.97 & 9.90 & 2.24 &  Irr                  \\
& CGCG097-125          &  11 44 54.97 &  19 46 24.7 &     15.74 &     8271  &  21.35 & 9.69 & 2.71 &  S0a                  \\
& NGC3861              &  11 45 4.33 &  19 58 21.0 &     13.47 &     5085  &  21.91 & 9.51 & 2.07 &  (R')SAB(r)b          \\
& NGC3862              &  11 45 5.23 &  19 36 37.8 &     13.67 &     6511  &  24.51 & 9.06 & -1.11 &  E;LERG               \\
& CGCG097-133NED01     &  11 45 17.32 &  20  1 17.3 &     15.71 &     5290  &  21.62 & 9.47 & 2.17 &  E                    \\
& CGCG127-049          &  11 45 49.50 &  20 37 29.6 &     15.40  &     7061  &  21.46 & 9.72 & 2.52 &  S-                   \\
& IC0732S              &  11 45 59.83 &  20 26 28.7 &     15.45 &     7311  &  22.24 & 10.50 & 2.49 &  S0/a                 \\
& NGC3884              &  11 46 12.23 &  20 23 31.0 &     13.50  &     6946  &  22.10 & 9.37 & 1.65 &  SA(r)0/a;LINER;Sy1   \\
\enddata
\end{deluxetable}

\begin{deluxetable}{llcccccccl}
\tablenum{2 Continued.}
\tablewidth{0pc}
\tablecaption{Cluster Galaxies in the Luminosity-Limited Sample}
\tablehead{
\colhead{}            &
\colhead{}               &
\colhead{$\alpha$}           &
\colhead{$\delta$}           &
\colhead{m}                  &
\colhead{$<cz>$}             &
\colhead{$\log L_{1.4GHz}$}   &
\colhead{$\log L_{60\mum}$}  &
\colhead{}                  &
\colhead{}              \\
\colhead{Cluster}                   &
\colhead{Name}                   &
\colhead{(2000.0)}           &
\colhead{(2000.0)}           &
\colhead{(mag)}              &
\colhead{(km~s$^{-1}$)}      &
\colhead{(W~Hz$^{-1}$)}      &
\colhead{(L$_{\odot}$)}      &
\colhead{q}                   &
\colhead{Notes}                   }
\startdata
Virgo & NGC4152              &  12 10 37.59 &  16  1 58.3 &     12.66 &     2167  &  20.90 & 8.96 & 2.32 &  SAB(rs)c;HII;Sbrst   \\
& MESSIER098           &  12 13 48.40 &  14 54 3.5 &     10.95 &     -142  &  21.26 & 9.24 & 2.30 &  SAB(s)ab;HII;Sy      \\
& NGC4208              &  12 15 38.69 &  13 54 1.3 &     11.83 &     -81   &  20.82 & 9.13 & 2.62 &  SAc                  \\
& MESSIER099           &  12 18 50.14 &  14 24 50.1 &     10.44 &     2407  &  22.02 & 9.87 & 2.16 &  SA(s)c               \\
& NGC4293              &  12 21 12.79 &  18 22 56.4 &     11.26 &     893   &  20.66 & 9.02 & 2.62 &  (R)SB(s)0/a;LINER    \\
& NGC4298              &  12 21 33.13 &  14 36 16.5 &     12.04 &     1135  &  21.03 & 9.01 & 2.37 &  SA(rs)c              \\
& NGC4302              &  12 21 42.36 &  14 35 44.3 &     12.50  &     1149  &  20.93 & 8.92 & 2.47 &  Sc:sp                \\
& MESSIER100           &  12 22 55.06 &  15 49 21.5 &     10.05 &     1571  &  21.92 & 9.74 & 2.13 &  SAB(s)bc;LINERHII    \\
& NGC4383              &  12 25 25.54 &  16 28 10.8 &     12.67 &     1710  &  20.85 & 9.26 & 2.61 &  Sa?                  \\
& NGC4388              &  12 25 46.97 &  12 39 43.7 &     11.76 &     2524  &  21.56 & 9.35 & 2.02 &  SA(s)b:sp;Sy2        \\
& NGC4402              &  12 26 7.71 &  13  6 50.3 &     12.55 &     232   &  21.11 & 9.06 & 2.30 &  Sb                   \\
& NGC4419              &  12 26 56.42 &  15  2 52.2 &     12.08 &     -261  &  21.09 & 9.21 & 2.38 &  SB(s)a;LINER;HII     \\
& NGC4438              &  12 27 45.56 &  13  0 31.0 &     11.02 &     71    &  21.46 & 8.96 & 1.83 &  SA(s)0/apec:LINER    \\
& MESSIER088           &  12 31 58.79 &  14 25 5.0 &     10.36 &     2281  &  21.84 & 9.57 & 2.12 &  SA(rs)b;Sy2          \\
& NGC4526              &  12 34 2.78 &  7  41 58.0 &     10.66 &     448   &  20.47 & 9.08 & 2.93 &  SAB(s)               \\
& NGC4535              &  12 34 19.90 &  8  11 50.7 &     10.59 &     1961  &  21.20 & 9.37 & 2.51 &  SAB(s)c              \\
& NGC4568              &  12 36 34.17 &  11 14 25.3 &     11.68 &     2255  &  21.53 & 9.64 & 2.43 &  SA(rs)bc             \\
& MESSIER090           &  12 36 49.62 &  13  9 56.9 &     10.26 &     -235  &  21.31 & 9.33 & 2.33 &  SAB(rs)ab;LINER;Sy   \\
& MESSIER058           &  12 37 43.74 &  11 49 8.0 &     10.48 &     1519  &  21.41 & 9.09 & 2.07 &  SAB(rs)b;LINER;Sy1.9 \\
& NGC4634              &  12 42 40.74 &  14 17 46.3 &     13.16 &     297   &  20.91 & 8.93 & 2.35 &  SBcd:sp              \\
& NGC4647              &  12 43 32.49 &  11 34 58.1 &     11.94 &     1422  &  21.13 & 9.05 & 2.28 &  SAB(rs)c             \\
& NGC4651              &  12 43 42.81 &  16 23 36.8 &     11.39 &     805   &  20.92 & 9.09 & 2.49 &  SA(rs)c;LINER        \\
& NGC4654              &  12 43 55.76 &  13  7 45.1 &     11.10  &     1037  &  21.48 & 9.47 & 2.30 &  SAB(rs)cd            \\
\\
Coma &  MRK0053              &  12 56 6.29 &  27 40 41.9 &     16.20  &     4968  &  21.67 & 9.74 & 2.36 &  Sa;HII               \\
& NGC4839              &  12 57 24.37 &  27 29 55.0 &     13.02 &     7362  &  22.88 & 8.99 & 0.40 &  cD;SA0               \\
& KUG1255+283          &  12 57 57.27 &  28  3 39.9 &     16.58 &     8299  &  21.68 & 9.50 & 2.03 &  Sb;Sbrst             \\
& NGC4848              &  12 58 5.44 &  28 14 37.2 &     14.41 &     7049  &  22.35 & 10.09 & 2.00 &  SBab:sp              \\
& CGCG160-058          &  12 58 10.06 &  28 42 32.6 &     15.48 &     7616  &  21.69 & 9.45 & 2.12 &  Sbc                  \\
& KUG1255+275          &  12 58 18.78 &  27 18 31.8 &     16.50  &     7389  &  21.72 & 9.56 & 2.11 &  Irr;Sbrst            \\
& NGC4853              &  12 58 35.30 &  27 35 42.4 &     14.41 &     7660  &  21.44 & 9.74 & 2.61 &  (R')SA0-?;AGN        \\
& MRK0056              &  12 58 36.00 &  27 16 4.5 &     16.20  &     7368  &  21.62 & 9.36 & 2.04 &  S0;Sbrst             \\
& MRK0057              &  12 58 37.57 &  27 10 33.2 &     15.40  &     7666  &  21.76 & 9.53 & 2.00 &  Irr                  \\
& MRK0058              &  12 59 5.06 &  27 38 34.8 &     15.12 &     5554  &  21.54 & 9.49 & 2.18 &  SBa;Sbrst            \\
& UCM1256+2722         &  12 59 16.12 &  27  6 13.8 &     17.32 &     8604  &  21.60 & 9.46 & 2.15 &  Sc+;Sbrst            \\
& NGC4869              &  12 59 21.53 &  27 54 38.9 &     14.77 &     6875  &  23.41 & 9.32 & 0.04 &  E3                   \\
& NGC4874              &  12 59 35.47 &  27 57 36.5 &     12.63 &     7224  &  23.32 & 9.12 & 0.04 &  cD;Di                \\
& RB038                &  12 59 37.41 &  27 59 17.2 &     15.70  &     6812  &  21.47 & 8.99 & 1.83 &  SA0/a                \\
& FOCA0430             &  12 59 39.57 &  28 10 31.9 &     18.61 &     5437  &  21.64 & 8.99 & 1.81 &  E                    \\
& KUG1258+279A         &  13  0 34.15 &  27 38 9.0 &     15.68 &     7476  &  21.64 & 9.28 & 1.95 &  Sb;Sbrst             \\
& KUG1258+278          &  13  0 35.63 &  27 34 40.9 &     15.50  &     5097  &  21.41 & 9.32 & 2.25 &  Sbc                  \\
& IC4040               &  13  0 37.97 &  28  3 23.2 &     15.36 &     7840  &  22.33 & 10.00 & 1.92 &  Sdm                  \\
& KUG1258+287          &  13  0 40.77 &  28 31 16.5 &     16.40  &     8901  &  21.81 & 9.24 & 1.59 &  Spiral               \\
& IC4041               &  13  0 40.07 &  28  0 14.2 &     15.30  &     7112  &  22.06 & 9.79 & 2.09 &  SA0                  \\
& NGC4911              &  13  0 55.81 &  27 47 27.7 &     13.55 &     7985  &  22.30 & 9.83 & 1.91 &  SAB(r)bc             \\
& KUG1259+289          &  13  1 24.61 &  28 40 48.9 &     15.31 &     8762  &  21.58 & 9.45 & 2.22 &  Irr                  \\
& NGC4927              &  13  1 57.92 &  28  0 18.1 &     14.95 &     7764  &  21.79 & 9.00 & 1.53 &  SA0-                 \\
& NGC4926A             &  13  2 7.48 &  27 38 55.9 &     15.22 &     7188  &  21.39 & 9.68 & 2.55 &  S0pec?;Sbrst         \\
& KUG1259+284          &  13  2 13.07 &  28 12 47.7 &     15.78 &     8202  &  21.44 & 9.40 & 2.22 &  Sb                   \\
& NGC4931              &  13  3 0.39 &  28  2 24.2 &     14.31 &     5443  &  21.53 & 8.92 & 1.71 &  SA0                  \\
\enddata
\end{deluxetable}

\begin{figure}[hbt]
\plotone{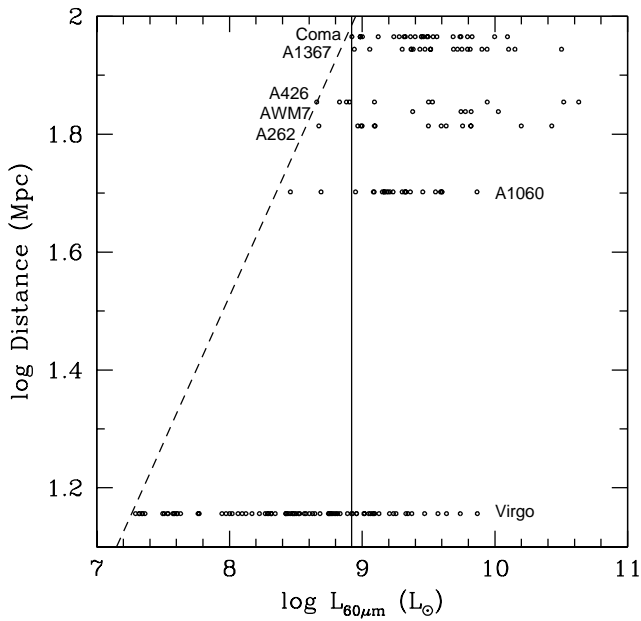}
\caption{Distribution of sources as a function of $60\mu$m luminosity.  Each
horizontal band is composed of all detected galaxies within a particular cluster.  The
dashed and solid lines denote the flux and luminosity limit ($\log L_{60\mum} = 8.92$), 
respectively, adopted for the sample.  Galaxies below the luminosity limit comprise
$37\%$ of the flux-limited sample, and most of these are in Virgo.\label{fig:lumvdist}}
\end{figure}

We have imposed a luminosity limit of $\log L_{60\mum} = 8.92$
(the $3\sigma$ IRAS detection limit for the most distant cluster in our sample, Coma, at
$92$ Mpc).  Figure~\ref{fig:lumvdist} shows the distances for the galaxies in  
the $7$ clusters as a function of $60\mu$m luminosity.  Over $37\%$ of the sample
is excluded by adopting our luminosity-limit, including the $10$ extended Virgo galaxies
for which accurate IRAS fluxes could not be found.  The final luminosity-limited catalog 
contains 114 sources.  The sample is small compared to optically-selected cluster 
samples, but is drawn from a homogeneous selection of clusters based
on X-ray emission and is luminosity-complete.  
The catalog of 114 galaxies is summarized in Table~\ref{tab:clgal}, 
along with the galaxies' blue magnitudes and RC3 morphological codes from NED.

A key motivation for this study is to understand the effect of ICM density on the star forming activity 
in member galaxies.  Using radius as a proxy for local density, each cluster has been divided into a Core and 
Ring region.  Core galaxies are defined as having a projected clustercentric distance
less than $0.5$ Mpc (see \S 3 for a justification of this radius),  
and Ring galaxies are defined to have a projected clustercentric distance 
between $0.5$ Mpc and $1.5$ Mpc.  We chose the maximum limit of $1.5$ Mpc since this is 
a median value for the virial radii of most clusters, typically between $1$ Mpc and $3$ Mpc.
Some fraction of Core galaxies are Ring galaxies appearing in projection
and the effect of this contamination is addressed in \S 3 and \S 4.3.

Morphological dependencies are investigated by dividing the sample into early-type
(E and S0) and late-type subsamples.  Of the $182$ galaxies, $54$ ($30\%$) are early-type 
galaxies, higher than the detection rate of early-type FIR galaxies 
in the field ($20\%$; see Cotton \& Condon 1998), and significantly higher than the fraction of 
ellipticals in the local universe ($\sim 15\%$).  This 
clustering of early-type galaxies reflects the density-morphology relation 
(Dressler 1980; \S 6.1).  It may also signify that cluster ellipticals are brighter at FIR wavelengths than 
those found in the field (\S 6.1).  The number of Core, Ring, early-, and late-type galaxies 
in the luminosity-limited sample for each cluster are summarized in Table~\ref{tab:clsum}.

\section{Deprojection and Density Distribution of Cluster Sources}

Accurate information on the nature of the Core galaxies can be obtained by 
estimating the amount of source contamination from the Ring region 
due to projection effects.  A geometrical recursive technique employing simple
discretization of the Abel integral is used to deproject 
the sources within spherical shells surrounding the core, starting with the
outermost shell.  The 
parameters that characterize the density distribution are the central density ($\rho_{\rm o}$), 
core radius ($r_c$),  and log-slope ($\beta$; see equations 4 and 5 below), and of these, 
$\rho_{\rm o}$ can be the most systematically uncertain: the outer shell has the lowest number count, 
and this statistical error 
can become significant after propagating to the inner shells, resulting in a range of normalizations.  
In contrast, $\beta$ and $r_c$ are more resilient to such initial
error as they parameterize a ``bulk'' behavior of the sources.  Minimizing the initial 
error in $\rho_{\rm o}$ requires accurate background decontamination, in which case velocity 
selection may be used to determine the most likely cluster members, as is done in this
study.  The cluster-to-cluster variation in density distribution is dominated by statistical 
noise, but there is a clear trend in that of the composite cluster.  Table~\ref{tab:cldeproj} 
lists the radii of the spherical shells, observed counts within those regions, computed 
deprojected counts, and estimated density, which increases with decreasing radius.  
The ``radius'' of a bin is defined as the mean of the inner and outer radii of that bin.  

In relaxed, regular clusters, the number density of galaxies and 
the gas density are correlated since the 
crossing time of a cluster galaxy is an order of magnitude shorter than a Hubble time if  
the cluster is in hydrostatic equilibrium.  
The hydrostatic model for these clusters provides a consistent fit for both the 
X-ray and number density distributions (Bahcall \& Lubin 1994).  
The galaxy surface density profile can be represented as
$$n = n_{\rm o} \left(1 + {r^2 \over r^{2}_{\rm c}}\right)^{(-3/2)\beta + 1/2}\eqno(4)$$
according to the $\beta$-model (Cavaliere \& Fusco-Femiano 1976).
The deprojected density that should fit both the number
and X-ray gas distribution is then given as:
$$\rho = \rho_{\rm o} \left(1 + {r^2 \over r^{2}_{\rm c}}\right)^{(-3/2)\beta}.\eqno(5)$$

\begin{figure}[hbt]
\plotone{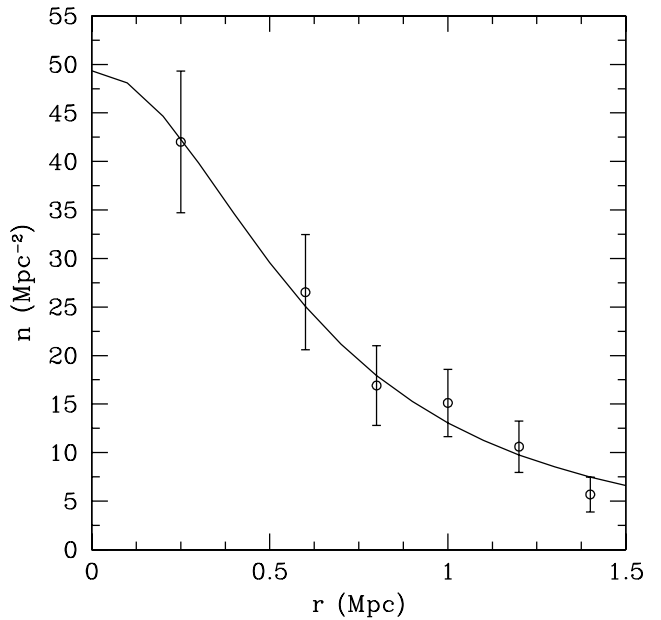}
\caption{Observed galaxy surface density (open circles) with best-fitting surface density
$\beta$ profile of equation (4) (solid curve).  Our best-fit values are $\beta \sim 1.05$, 
$r_{\rm c} \sim 0.64$ Mpc, and $n_{\rm o} \sim 49$ Mpc$^{-2}$, with $\chi^{2} = 2.09$.
The density drops by a factor of $\sim 10$ from $0$~Mpc to $1.5$~Mpc.
\label{fig:surfacefit}}
\end{figure}

\begin{deluxetable}{lcccc}
\tablewidth{0pc}
\tablecaption{Deprojection Statistics\label{tab:cldeproj}}
\tablehead{
\colhead{}                  &
\colhead{Radial Boundaries}       &
\colhead{}         &
\colhead{}      &
\colhead{Number Density}          \\
\colhead{Region}                        &
\colhead{(Mpc)}                     &
\colhead{Observed Counts}                        &
\colhead{Deprojected Counts}                        &
\colhead{(Mpc$^{-3}$)}}
\startdata
0 (Core) & $0.0<r<0.5$ & 33 & 15.92 & 30.41 \\
1 & $0.5<r<0.7$ & 20 & 17.85 & 19.55 \\
2 & $0.7<r<0.9$ & 17 & 10.16 &  6.28 \\
3 & $0.9<r<1.1$ & 18 & 19.49 &  7.73 \\
4 & $1.1<r<1.3$ & 16 & 22.47 &  6.19 \\
5 & $1.3<r<1.5$ & 10 & 28.11 &  5.70 \\  
Total &  ...  & 114 & 114.00 &  ...  \\
\enddata
\end{deluxetable}

The accuracy of the deprojection method can be determined by fitting the galaxy surface density
with equation (4) (Figure~\ref{fig:surfacefit}).  While the $95\%$ confidence limits for
$n_{\rm o}$, $r_{\rm c}$, and $\beta$ are a factor of two times the best-fit values, the
mean values of $n_{\rm o}$ and $r_{\rm c}$ are consistent with those found for typical clusters  
($n_{\rm o} = 49$ Mpc$^{-2}$, $r_{\rm c} = 0.64$ Mpc, and $\beta = 1.05$ 
with $\chi^2 = 2.09$).  Comparison with the independent fit for the deprojected galaxy density 
(not shown) yields parameters that agree with those in Figure~\ref{fig:surfacefit}.  These
results indicate that selection based on radio and FIR emission does not bias the galaxy
distribution away from that seen among clusters in general.

\section{FIR-Radio Relation}

\subsection{Radio-FIR Correlation}

\begin{figure}
\centerline{\epsfxsize=8cm\plotone{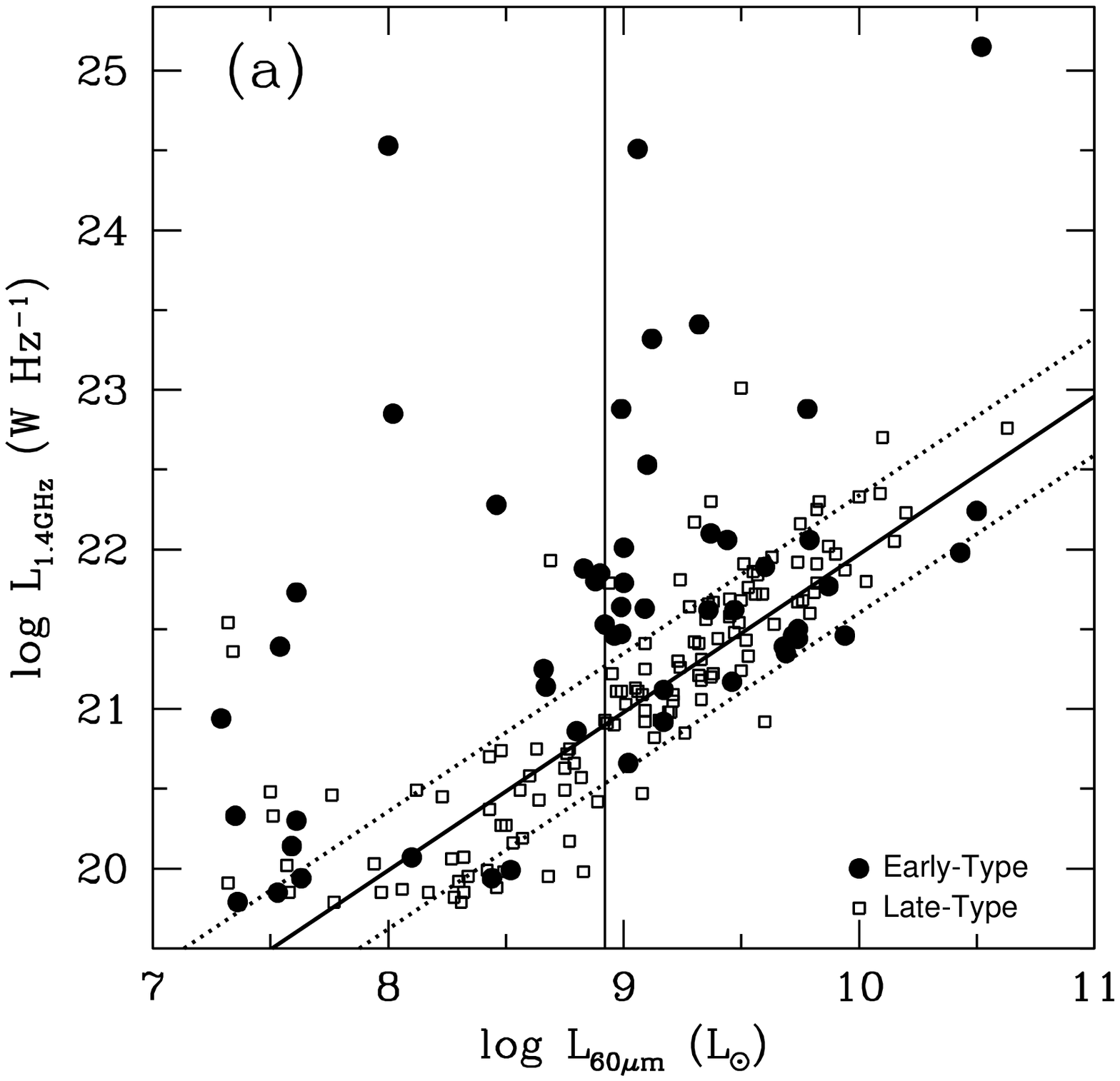}}
\centerline{\epsfxsize=8cm\plotone{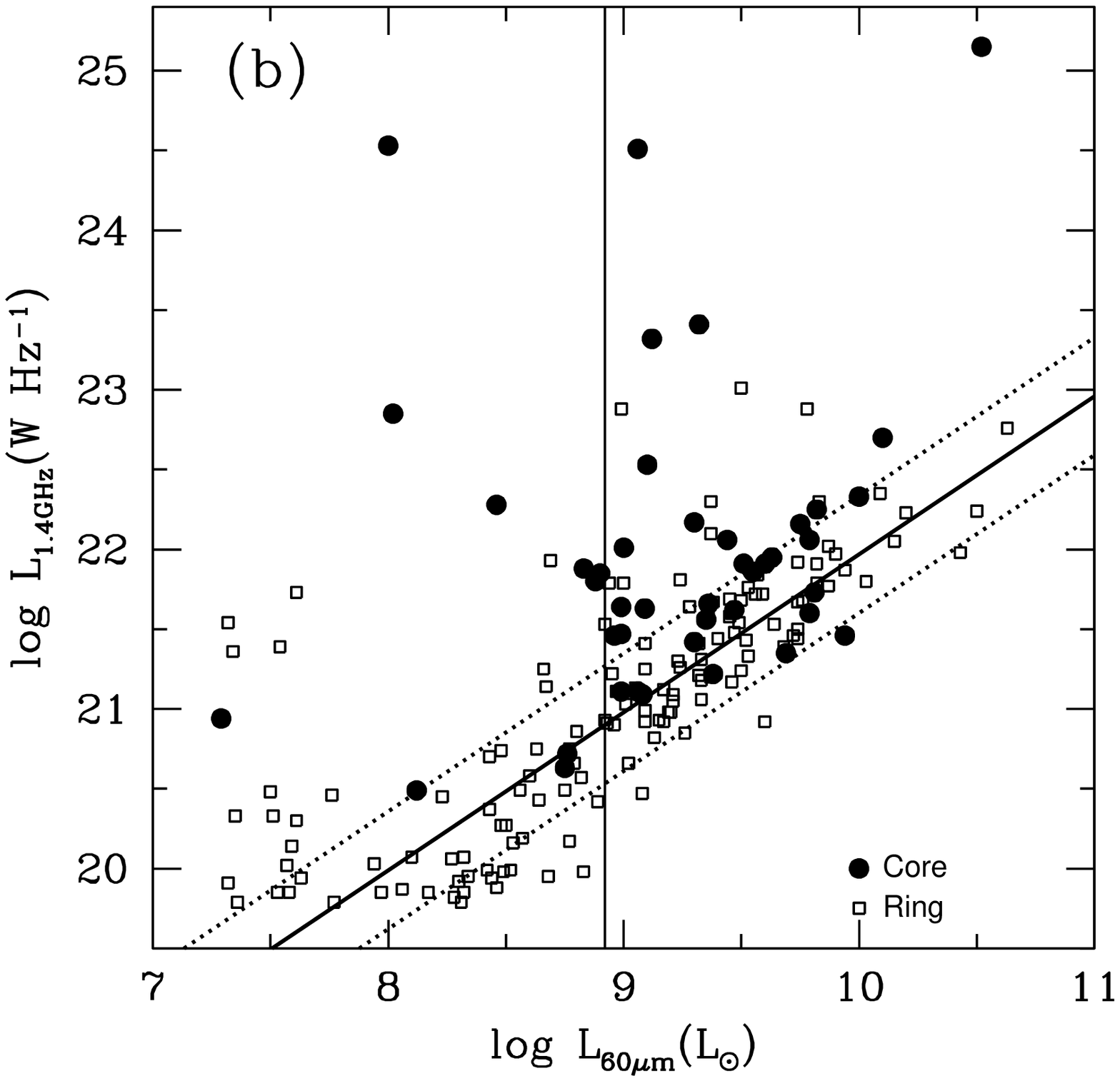}}
\caption{Plot of $20$ cm continuum radio emission (eq. [1]) versus
IRAS $60\mu$m luminosity (eq. [2]) for the flux-limited sample.  The sloping solid line denotes 
the best-fit linear correlation for the YRC field sample and the two dotted lines 
indicates its 1$\sigma$ dispersion.  The vertical line delineates our
adopted luminosity-limit.  Large circles identify early-type and Core galaxies in Figure~\ref{fig:lums}a and 
Figure~\ref{fig:lums}b, respectively, which are biased above the field radio-FIR correlation.
The small points refer to late-type and Ring galaxies in Figure~\ref{fig:lums}a and 
Figure~\ref{fig:lums}b, respectively.  Note the significant number of sources above
the correlation compared with the few that fall below.
\label{fig:lums}}
\end{figure}

Figure~\ref{fig:lums} shows the $20$ cm radio continuum luminosity plotted 
against the $60\mu$m luminosity, i.e., the radio-FIR
correlation, for the flux-limited sample.  The band between the two dotted lines
indicates the best-fit linear correlation between radio and FIR luminosity and the
scatter of $0.26$ dex for the 1809 galaxies of the IRAS 2 Jy YRC field sample.  
Most ($98\%$) of the IRAS selected field
galaxies, including early-types (E and S0), are normal star forming galaxies following
the linear radio-FIR correlation (YRC).  Comparison with these field galaxies
shows that $\sim 37\%$ of the cluster galaxies deviate significantly ($>0.26$ dex) from the
radio-FIR correlation defined by the IRAS-selected sample.  For comparison, $<2\%$ of field
galaxies deviate $>0.26$ dex from the mean relation, mostly from systematic deviations
at the high and low luminosity ends.  Twenty-seven percent
of the cluster galaxies above our flux limit exhibit a higher radio luminosity than expected
from the linear field correlation ($> 1\sigma$ above the field correlation): 
$50\%$ of these galaxies are in the {\it observed} Core region and $62\%$ are early-type galaxies.

None of the sources have $L_{60\mum}\geq 10^{11} L_{\odot}$, in contrast to the field where $\sim 11\%$ are
luminous in the infrared (YRC).  The 2 Jy selection at $60\mum$ is identical to the optical
selection of late-type galaxies in the field (YRC).  We expect the same selection effect for
our cluster sample.  Rarity of IR luminous galaxies has been noted
in other similarly-sized samples of cluster galaxies (e.g, Bicay \& Giovanelli 1987).  From
YRC, the space density of IR luminous galaxies in the field is about $10^{-5}$ 
Mpc$^{-3}$, and there is only a $0.1\%$ chance of finding such a source within
the search volume ($\sim 100$ Mpc$^{3}$) in the field.  The high galaxy density in the 
cluster environment should increase the expected fraction by an order of magnitude to
near unity.  Their non-detection in these clusters by itself is not a strong indication
of bias against IR luminous phenomena in the cluster environment.
For comparison, the Abell cluster study of Miller \& Owen (2001b) finds only 4 of 329 cluster
sources with secure IRAS $60\mu$m and $100\mu$m detections having $L_{\rm FIR} > 10^{11}$ 
L$_{\odot}$, and only $1$ of these $4$ lies $>1.5$~Mpc from the cluster center.

A separate linear least-squares fit to the radio and FIR luminosity for our 182
sample galaxies indicates a shallower than unity log-slope of $0.76$, but with a
dispersion of 0.74, emphasizing the much greater scatter among the cluster galaxies
as compared with those from the field (Figure~\ref{fig:lums}).
The scatter is slightly less if we only consider galaxies in the luminosity-limited sample.  
The shallower slope is not statistically significant given the large scatter among
cluster galaxies.

Note the apparent bias of sources below the best-fit
correlation between $\log L_{60\mum}=8.00$ and our luminosity cutoff of $\log L_{60\mum}=8.92$.  
Approximately $90\%$ of the galaxies below the luminosity cutoff are in Virgo.  The
excess of radio-luminous galaxies at $\log L_{60\mum} = 7.5$ L$_{\odot}$ consists of
mostly elliptical and/or Seyfert ($9/16$) galaxies, none of which are included in the Virgo
radio-FIR study of Niklas, Klein, \& Wielebinski (1995).  One source (IC3099) in this region 
is one of the $10$ for which we could not find accurate IRAS fluxes: the radio bias of this 
source may be due to missing extended FIR emission.  Niklas et al. (1995) identify a few Virgo 
early-type galaxies in the Core with excess $2.8$ and 
$6.3$ cm radio emission compared with their FIR emission.  This is also the case for early-type
galaxies further away from the cluster center (Figure~\ref{fig:lums}).
Excluding these radio-luminous objects, the majority of the remaining galaxies 
are situated below the radio-FIR correlation for the YRC field sample, and a formal fit to the
data yields a greater than unity slope, consistent with the Niklas et al. (1995) study and
with the nonlinearity seen at lower luminosities in the radio-FIR correlation of the YRC field
sample (Fig. 5a of YRC; see also Bell 2002 for a more recent review).  
Possible explanations for this nonlinearity include dust heating by low-mass stars or 
cosmic-ray diffusion.  
Adopting a luminosity cutoff excludes this nonlinear regime and selects the region in 
luminosity-luminosity space where both the FIR and radio luminosity are directly proportional 
to the massive star formation rate, allowing for an unbiased method of segregating star forming galaxies 
and AGNs (see \S 6.3).

\begin{figure}
\centerline{\epsfxsize=8cm\plotone{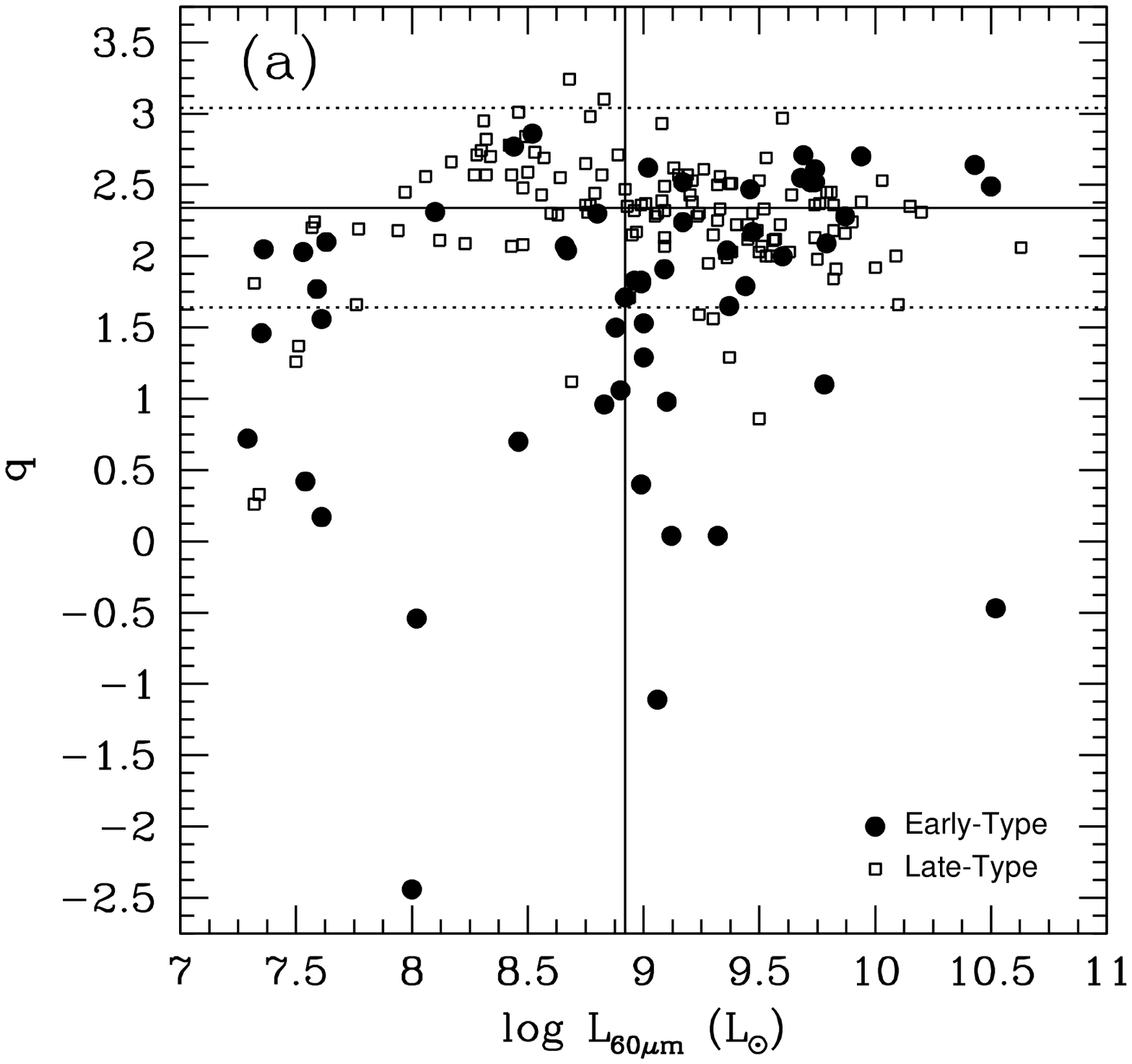}}
\centerline{\epsfxsize=8cm\plotone{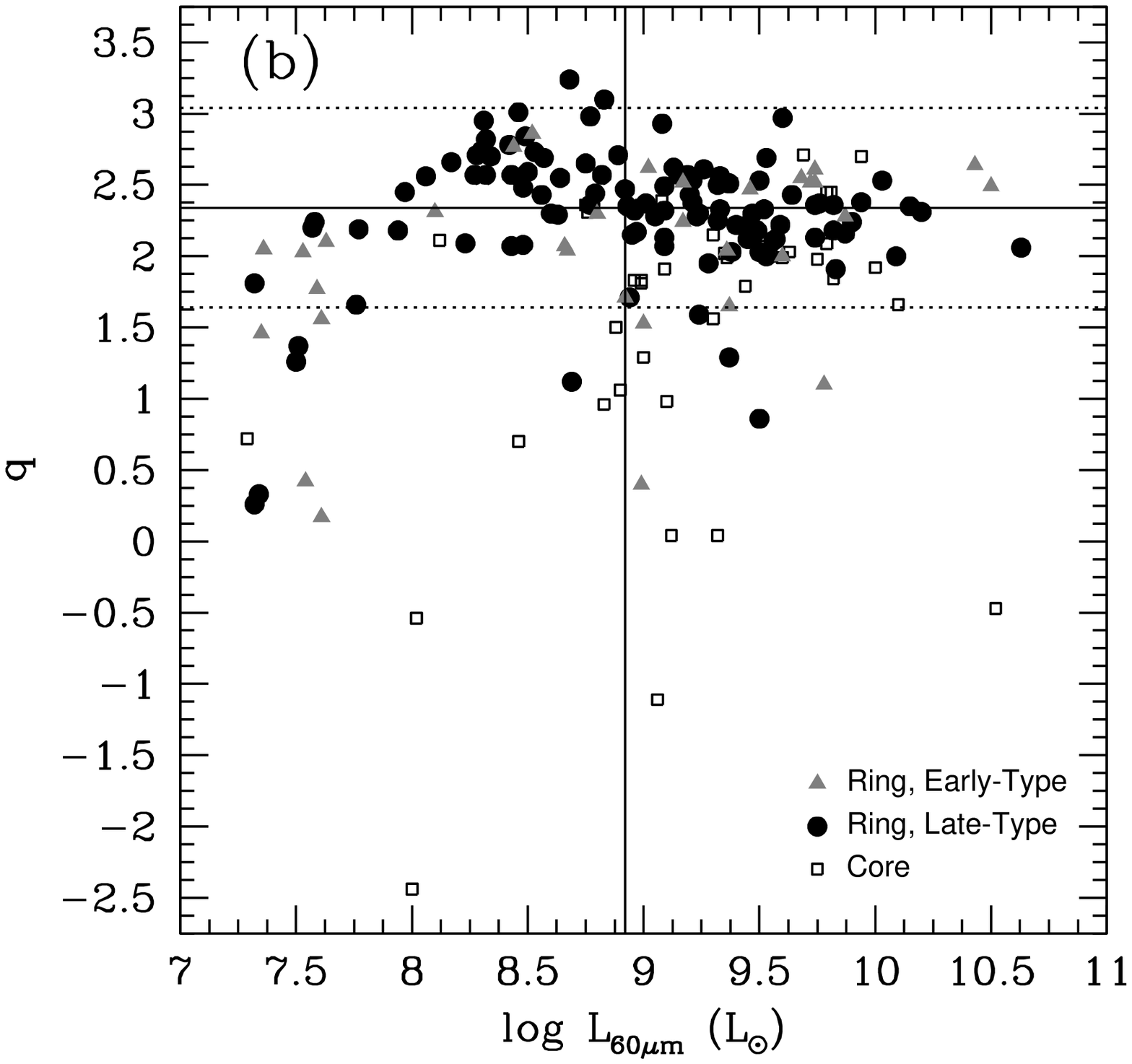}}
\caption{Logarithmic FIR-radio ratio, $q$ parameter, plotted for each galaxy
as a function of IRAS $60\mu$m luminosity, with separate emphasis on morphological and
radial dependence in Figure~\ref{fig:qplot}a and Figure~\ref{fig:qplot}b, respectively.  
Filled circles and open squares represent early- and
late-type galaxies, respectively, in Figure~\ref{fig:qplot}a.  
Filled grey triangles and circles represent early- and late-type Ring
galaxies, respectively, and open squares denote Core galaxies in Figure~\ref{fig:qplot}b.
The solid horizontal line denotes the average $\langle q\rangle = 2.34$ found for YRC
field galaxies, dotted lines delineate IR- and 
radio-excess regions, and the solid vertical line denotes our luminosity-limit of 
$\log L_{60\mum} = 8.92$.\label{fig:qplot}}
\end{figure}

\clearpage

\subsection{FIR-Radio Flux Density Ratio}

To investigate the deviation from the linear radio-FIR correlation, we compute 
the $q$ parameter (Condon, Anderson, \& Helou 1991),
$$q \equiv \log \left[{{FIR}\over{3.75 \times 10^{12}~{\rm W~m}^{-2}}}\right] - \log \left[{{S_{1.
4 \rm GHz}}\over{{\rm W~m}^{-2} {\rm Hz}^{-1}}}\right]\eqno(6)$$
where $S_{1.4 \rm GHz}$ is the observed 1.4 GHz flux density in units of 
W m$^{-2}$ Hz$^{-1}$ and
$$FIR \equiv 1.26 \times 10^{-14}(2.58S_{60\mum}+S_{100\mum})~{\rm W~m}^{-2} ,
\eqno(7)$$
where $S_{60\mum}$ and $S_{100\mum}$ are IRAS 60 $\mu$m and 100
$\mu$m band flux densities in Jy (see Helou et al. 1988).
The $q$ parameter is both independent of distance and starburst strength (YRC; Lisenfeld,
Volk, \& Xu 1996) and so is a good quantitative measure for comparing galaxies independent
of their star formation rates.  The $q$ values for the cluster galaxies are listed in 
Table~\ref{tab:clgal} and are plotted in Figure~\ref{fig:qplot}.  The lines appearing at $q=3.04$ and $q=1.64$ delineate
the point at which the galaxies have $5$ times larger IR and radio flux, respectively, than what
would be expected from the radio-FIR correlation.  Sources with $q>3.04$ and $q<1.64$ are 
referred to as IR- and radio-excess sources, respectively (YRC).

The mean $q$ for the flux-limited sample is $2.07\pm 0.74$, 
{\it statistically} consistent with the mean value of $q=2.34\pm 0.01$ for the field.  
However, the low $q$ value and large scatter as compared with the field is due in part 
to AGN activity, which results in a {\it systematic} difference in the $q$ distribution.  
The mean $q$ for the luminosity-limited sample is similar, $\langle q\rangle =2.10$, with smaller dispersion, despite the 
exclusion of low radio luminosity sources among the low IR luminosity population of Virgo 
discussed above.  This indicates that the radio-luminous sources below our luminosity
cutoff (e.g., elliptical and Seyfert galaxies in Virgo) balance the effect of the
galaxies appearing below the radio-FIR correlation.  About $1/2$ of the Abell
cluster galaxies of Miller \& Owen (2001a) populate poor clusters and are projected
up to $3$ Mpc from the cluster center, yet have a lower mean $\langle q\rangle=1.95$
($\sigma = 0.77$).  This is likely due to their selection based strictly on radio emission
and the presence of a large number of optically-identified AGNs in their sample.

\subsubsection{cD Galaxies}

There are a total of $13$ radio-excess sources in our sample, $4$ of which are cD galaxies
(NGC708, NGC1275, NGC4839, NGC4874).  Radio and
X-ray observations of these cD galaxies imply jet-like activity in their nuclei 
(Conselice, Gallagher, \& Wyse 2001;  Holtzman et al. 1992;  Vermeulen, Readhead, \&
Backer 1994; Rhee, Burns, \& Kowalski 1994; Neumann et al. 2001; Colles \& Dunn 1996; Feretti
\& Giovannini 1987).  The FIR luminosity may result from 
star formation, but in some cases, such as NGC1275, there is also a notable contribution from
collisional heating of dust (e.g., Irwin, Stil, \& Bridges 2001---see \S 6.4.1 for further
discussion).

\subsubsection{Remaining Radio-Excess Sources}

Of the remaining $9$ radio-excess sources, $5$ show radio, X-ray, and/or emission lines
consistent with the presence of an AGN. (Bonatto et al. 1996; Owen, Ledlow, \& Keel 1996; 
Sambruna, Eracleous, \& Mushotzky 1999; V$\acute{e}$ron-Cetty \& V$\acute{e}$ron 1996; 
Ji et al. 2000; Baum et al. 1997; de Koff et al. 2000; Bai \& Lee 2001; Zirbel \& Baum 1995).  
CGCG097-073 is a spiral galaxy in A1367 that shows strong and extended H$\alpha$
emission, indicating vigorous star formation activity (Moss, Whittle, \& Pesce 1998). 
KUG1258+287 is an HI rich spiral galaxy in Coma with relatively low optical surface brightness in
the Digital Sky Survey image and may be interacting with its close companion FOCA195
(Bravo-Alfaro et al. 2000).  NGC4927 is another early-type Coma galaxy.  
NGC3884 is close to the boundary for a radio-excess galaxy, with $q=1.65$.
It is a Seyfert galaxy, with an extended LINER spectrum (Durret et al. 1993).  

Thirteen of $114$ galaxies ($11\%$)
show radio-excess, $11$ times higher than the detection rate in 
the YRC field sample.  Of the $13$ radio-excess sources,
$9$ are early-type galaxies.  Of the $15$ and $18$ early-type galaxies in the Core and
Ring region (in the luminosity-limited sample),  $40\%$ and $17\%$, respectively,
are radio-excess sources.  As discussed above, at least $5$ of the non-cD radio-excess sources
host an AGN.  The average $q$ for both late- and early-type observed
Core galaxies is $1.74$, a FIR-radio flux density ratio half that of the cluster
average.  The average $q$ in the Ring is $2.24$, still a factor
of $1.3$ times higher than the mean $q$ for the YRC field sample.  
These results suggest some variation in $q$ with radial distance.

\subsection{$q$-Radius Correlation}

\begin{figure}[hbt]
\plotone{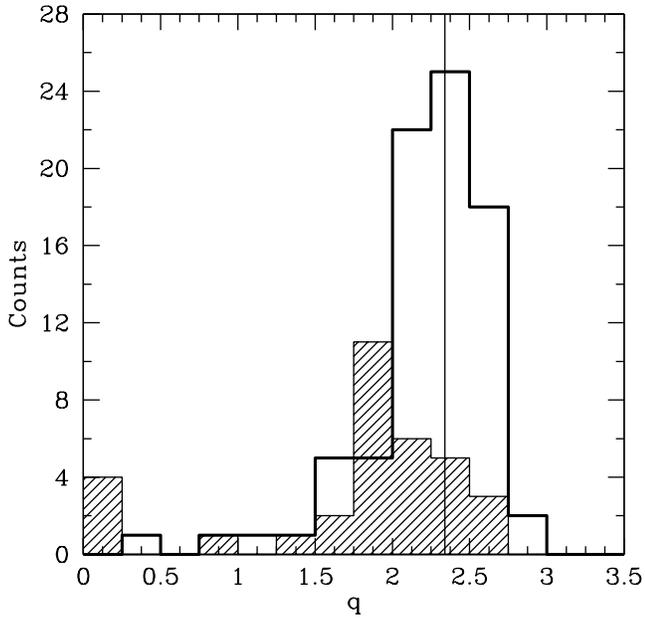}
\caption{Distribution of $q$ values for the luminosity-limited sample of cluster galaxies.  Sources
with $q<0$ are included in the first bin.  The hashed and unhashed portions of the distribution 
are for {\it observed} Core and Ring galaxies, respectively.  The solid vertical line delineates the average 
$q$ value for the YRC field sample.  Deprojection analysis suggests that most of 
those Core galaxies with $q>2.00$ are Ring galaxies seen in projection (see text).
\label{fig:qhist}}
\end{figure}

The distribution of galaxies as a function of their $q$ values is shown in 
Figure~\ref{fig:qhist} (binned every 0.25 dex).  Objects with negative
$q$ values are included in the first bin.  The distribution for
Ring galaxies peaks near the mean $q$ value for the YRC field sample, whereas \
the peak for the Core galaxies lies almost a factor of $2.3$ 
lower.  Both distributions drop off quickly for high $q$ values, and taper off 
more slowly for lower $q$ values, indicating a general bias toward lower $q$ values.

The deprojection analysis (Table~\ref{tab:cldeproj}) indicates that $17$ of the 
observed $33$ Core members could be Ring galaxies seen in projection.  Assuming
these contaminating sources have the same $q$ distribution
as the Ring sources, then $95\%$ of
the observed Core galaxies with $q>2.00$ are Ring members, and will be removed after 
deprojection and ``added'' to the Ring sample.  If this statistical result reflects a 
real change in the population, then the average $q$
value of the Core galaxies drops from $1.74$ to $1.18$, a factor of $14$ decrease
from that of field galaxies (YRC).

Of the Ring galaxies, $83\%$ have $q>2.00$.
Similarly, $91\%$ of Core galaxies have $q<2.00$.  If we assume that all galaxies
with $q<2.00$ are Core galaxies and all with $q>2.00$ are Ring galaxies, then
``deprojecting'' according to $q$ value indicates that $14$ and $95$ galaxies are
in the Core and Ring regions, respectively.  For comparison, deprojection
based on the number counts yields $17$ and $99$ Core and Ring sources, respectively.
If the above assumptions are made, then the $q$ value can be used to roughly discriminate 
between Core and Ring galaxies, and would be useful in 
situations where the density profile of cluster galaxies cannot be determined accurately.  
A more meaningful segregator may be observed with a larger sample of data and should
also vary with core radius and cluster richness.  A detailed investigation is necessary 
to establish the strength of, and variation in, the $q$ segregator.

According to Figure~\ref{fig:qplot}, $7$ of $33$ ($21\%$) observed Core galaxies
have radio-excesses, and $6$ of these are early-type galaxies.  Similarly,
$6$ of $81$ ($7\%$) Ring galaxies have radio-excesses, and $2$ of these are
early-type galaxies.  The increase in frequency
of radio-excess sources toward the center of clusters contributes to a radius-dependent 
$q$ distribution.  There is significant scatter in the correlation
between $q$ and $r$, and the best fit relation is
$$q = (0.43\pm 0.12) \times r + (1.84\pm 0.10),\eqno(8)$$  
where the 4 cD galaxies in our sample have been
excluded for this analysis---all of which are classified as radio-excess 
sources based on our definitions---since these are not typical of cluster galaxies.  

If all radio-excess sources are excluded, the correlation between $q$ and $r$ is shallower 
with a scatter of 0.28 dex, identical to the scatter in $q$ for YRC 
field galaxies.  Consequently, the scatter does not reflect statistical noise, but is a 
systematic result of the radio-excess sources in our sample, most of which are AGNs.  
If the increased radio luminosity of elliptical galaxies stems from  
an increase in number density, then the correlation of equation (8) reflects the 
density-morphology relation as the early-type galaxies have a density 
distribution that peaks in the Core region.  

\begin{figure}
\centerline{\epsfxsize=8cm\plotone{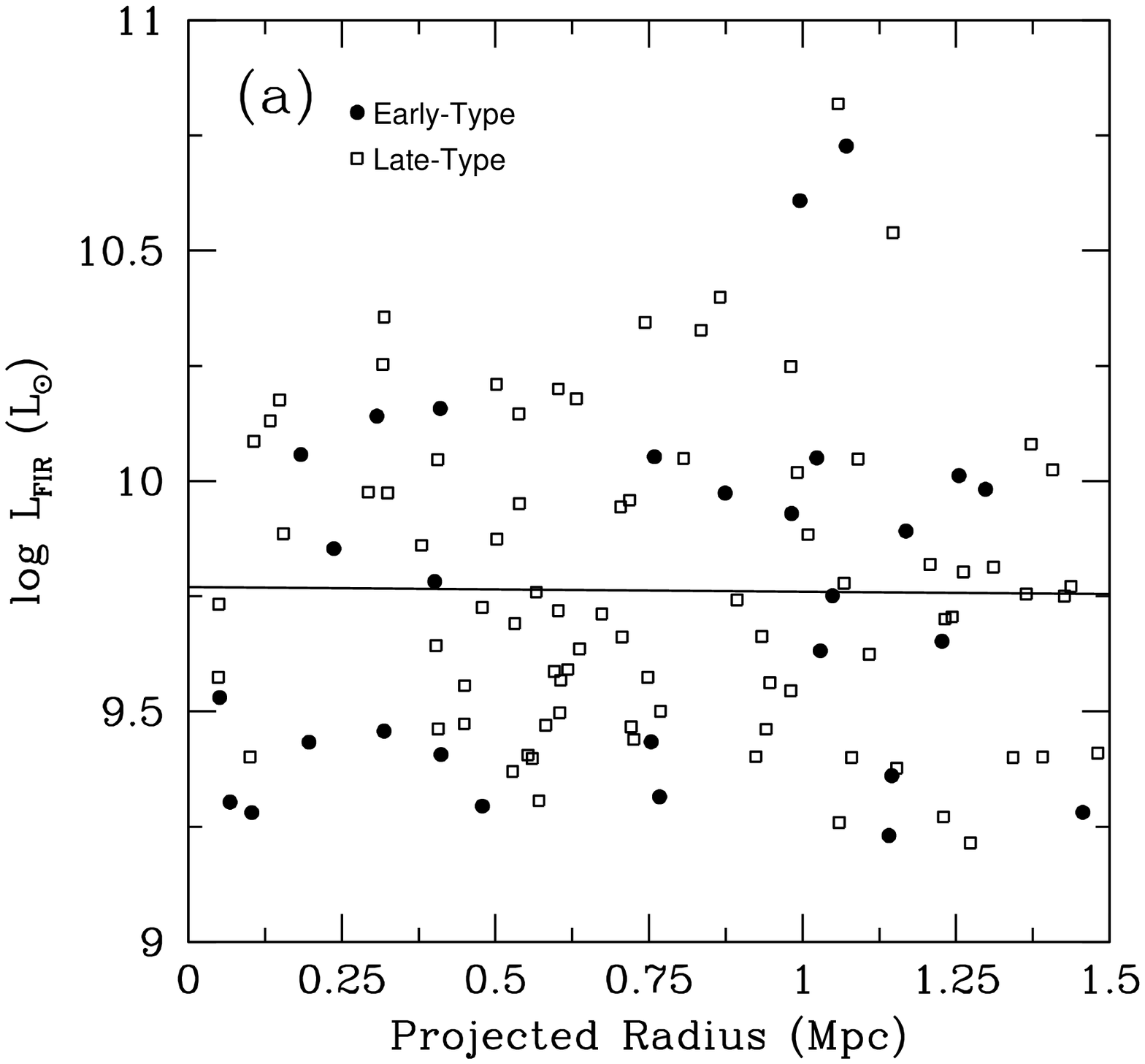}}
\centerline{\epsfxsize=8cm\plotone{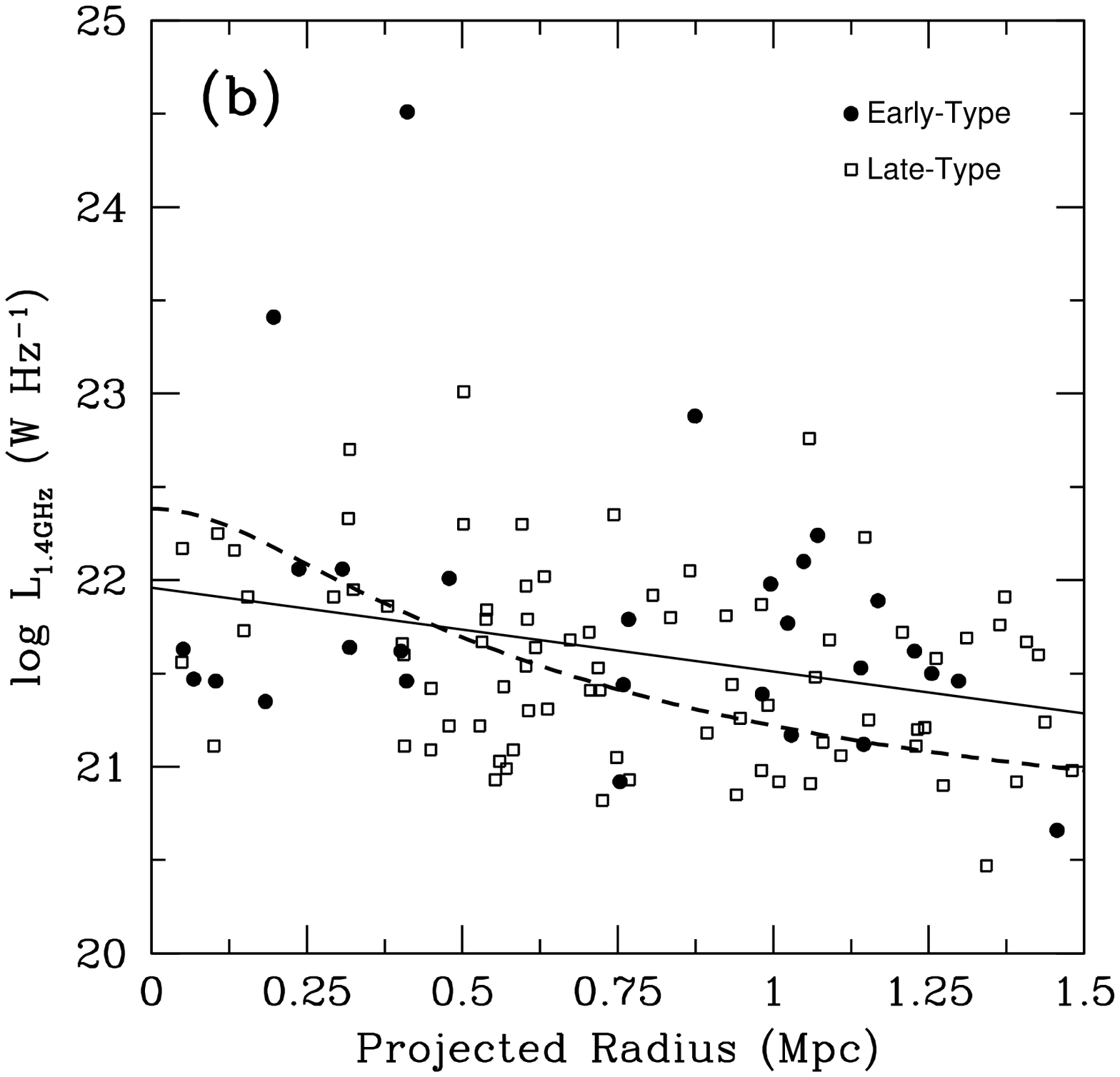}}
\caption{$\log L_{60\mum}$ and $\log L_{1.4\rm GHz}$ luminosity as a function of
radius for the luminosity-limited sample 
in Figure~\ref{fig:lumvrad}a and Figure~\ref{fig:lumvrad}b, respectively.  
The solid lines indicate the best-fit correlations.  The dashed curve
indicates the expected radio luminosity as a function of radius for a field spiral
galaxy undergoing thermal compression by the ICM (see \S 6.5.2).  Solid circles are 
early-type galaxies and open squares are late-type galaxies.  The correlation for 
the IRAS $60\mu$m luminosity is flat, whereas the radio luminosity 
shows some variation with radius.  The correlation between $q$ and $r$
can be completely accounted for by the correlation between $\log L_{1.4\rm GHz}$
and $r$.\label{fig:lumvrad}}
\end{figure}

Figure~\ref{fig:lumvrad} shows the radio and FIR luminosity (computed from eq. [3]) for the
luminosity-limited sample (excluding cD galaxies) as a function of radius.  Formal fits to the 
correlations give $\log L_{1.4 \rm GHz} \propto (-0.45\pm 0.14)\times r$ and
$\log L_{60\mum} \propto (-0.01\pm 0.09)\times r$, giving $q \propto (-0.01+0.45)\times r = 
(0.44\pm 0.23)\times r$,
and indicates the general bias of radio luminous sources 
toward the Core region.  We also note that the fraction of galaxies with
$\log L_{1.4\rm GHz}<21.00$ increases from $0\%$ in the Core to $18\%$ in the Ring.  
FIR luminosity is constant with radius.  The lack of FIR luminosity segregation has 
also been noted by Rengarajan, Karnik, \& Iyengar (1997) for spiral cluster galaxies.  We find
the same to be true for early-type galaxies.  Hence, $q$ values are lower in the 
Core than the Ring region due primarily to an increase in the radio luminosity, as opposed 
to a deficit in the FIR luminosity.  Possible variations in the FIR luminosity-radius 
correlation between clusters are investigated in \S 6.4.2.

\section{Radio and IR Cluster Luminosity Functions}

Comparison of the cluster galaxy luminosity distribution with an appropriately chosen
field sample should draw out distinctions between the cluster and field
populations, and we can thereby investigate whether the rich ICM preferentially 
affects the faint or bright ends of the cluster galaxy population.  We do this analysis by
computing the cluster radio and IR luminosity functions (LFs)
via the $1/V$ method (Schmidt 1968; Felten 1976)---i.e., number of sources per luminosity interval
divided by the total volume equal to the number of clusters falling below the
luminosity interval times the volume of one cluster.  For simplicity, we have assumed the
volume of a single cluster to be that of a sphere of radius $1.5$ Mpc.  In practice, the
sampled volume (and hence normalization of the LF) may be difficult to constrain due to a 
number of factors (e.g., projection effects, the adoption of a $2000$ km~sec$^{-1}$ velocity 
cutoff, etc.).  The deprojection results of \S 3 suggest that the assumption of a 
spherical volume is not unreasonable.  Furthermore, our adopted velocity cutoff should only
affect the normalization of the LF since all types of galaxies (e.g., Core, Ring, early-, and late-type)
are found to uniformly populate the range of luminosities considered here.  We are most interested 
in comparing the {\it shapes} of the cluster and field LFs and not the exact normalization.  Finally, a joint analysis 
of the radio and IR LFs should mitigate some of the uncertainties introduced by sample
selection and volume determination.

The LF is constructed by first computing the number of galaxies per luminosity
interval, taken to be $\Delta \log L_{60\mum} = 0.25$ for 
$\log L_{60\mum}$ between $7.50$ and $11.00$.  
The cluster LF is compared to the best-fit Schechter (1976) LF for the low-luminosity YRC field 
population, normalized to the value of the LF in the luminosity 
bin containing the greatest number of sources (representative of $L^*$ galaxies).    

\subsection{IRAS $60\mu$m Cluster Luminosity Function}

\begin{figure}
\plotone{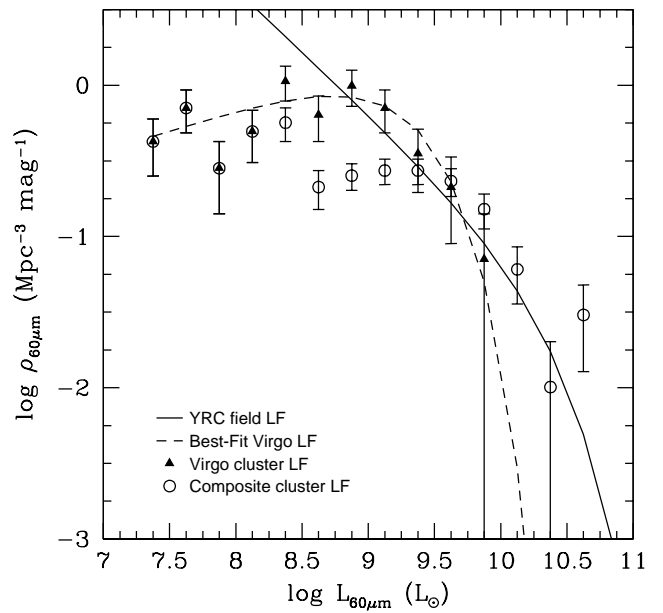}
\caption{The IRAS $60\mu$m LF for the 182 sources in the composite cluster is shown
by the open circles.  The solid curve denotes the best-fit low luminosity 
field Schechter LF for the $60\mu$m data of the YRC field sample,
normalized to the value of the cluster LF for the bin containing the greatest 
number of sources.  The systematic bias of the cluster LF below that of the best-fit 
field correlation suggests incompleteness at the low luminosity end.  For comparison, 
we have also shown the IR LF for Virgo cluster sources only, denoted by the filled 
triangles, along with its best-fit luminosity function indicated by the dashed curve.
\label{fig:lf60}}
\end{figure}

The cluster LF is well-described by the shape of the low-luminosity field LF (Figure~\ref{fig:lf60};
YRC) for $\log L_{60\mum} > 9.0$, after renormalizing the field LF by a factor of $\sim 180$.  
The break in the LF at $\log L_{60\mum} \sim 8.5$ is a result of the seven clusters not being uniformly 
distributed in redshift space, where the sampled volume increased by a factor of $2$.  For comparison, 
the LF computed for Virgo cluster galaxies is also shown in Figure~\ref{fig:lf60}.
The paucity of IR and radio sources in the Virgo core (\S 6.4.2) indicates that the sampled volume for
Virgo may be smaller than the one taken here, yet correcting for this does not fully account for
the disparity between the cluster and field LFs at low luminosity.
The bias of the cluster LF below the field LF may therefore be a 
result of incompleteness resulting from non-detections of low luminosity Virgo sources.  
Previous studies indicate $\alpha \sim -1.3$ for optically-selected Virgo galaxies (e.g., Ferguson
\& Sandage 1988).  Assuming that the FIR emission from faint blue galaxies is correlated with 
their optical luminosity, the faint-end slope for the Virgo cluster LF in this study is inconsistent
with the Ferguson \& Sandage (1988) result.  The divergent behavior of the low-luminosity end
of the IR LF is discussed below (\S 5.3).

\subsection{$1.4$ GHz Cluster Radio Luminosity Function}

\begin{figure}
\plotone{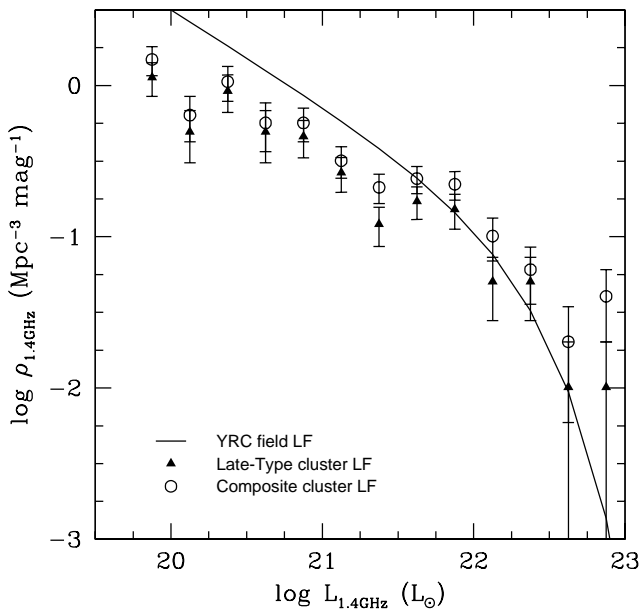}
\caption{The $1.4$ GHz radio luminosity function for the 182 sources in the composite
cluster is shown by the open circles, and that for the late-type cluster 
galaxies is shown by the filled triangles.  The solid curve denotes the best-fit 
field Schechter LF for the radio data of the YRC field sample,
normalized to the value of the cluster LF for the bin containing the greatest number of
sources.  We find the radio LF to more closely match that of the field sample. 
Three sources with $\log L_{1.4\rm GHz} > 24.5$ are not shown here, and they contribute
to the systematic excess of sources above the field LF for high luminosity objects,
most of which is due to the radio luminous early-type galaxies.
\label{fig:lf14}}
\end{figure}

Figure~\ref{fig:lf14} shows the cluster radio LF, constructed in the same way
as the IR LF.  The cluster radio LF is scattered around the field LF.  
Unlike the situation with the IR LF, the low-luminosity cluster radio sources appear to
have a faint-end slope similar to that of the field.  This can be easily explained
by the increased contribution from radio-luminous Virgo sources with
$\log L_{60\mum}$ between $7$ and $8$ (e.g., Figure~\ref{fig:lums}). 
There is an excess of cluster galaxies at the high luminosity end which is
more pronounced than that seen in Figure~\ref{fig:lf14}, as there are 3 sources with
$\log L_{1.4\rm GHz} > 24.5$ that are not shown in the figure.  The sources 
showing excess radio emission are seen more clearly in Figure~\ref{fig:qplot}, where the 
scatter of cluster galaxies is not symmetric about the scatter among 
YRC field galaxies. 
A similar excess is found in the radio LF for YRC galaxies, and after removing all 
sources within $1.5$ Mpc of the centers of known clusters within the limits of the YRC 
field sample, the field galaxies have a LF indistinguishable from the UGC sample LF of the 
Condon (1989) study.  The excess observed in Figure~\ref{fig:lf14}, as well as the excesses
observed for the IRAS~2 Jy sample (see Figure~8b in YRC) and the radio LF of UGC sample 
galaxies (Condon, Cotton, \& Broderick 2002) all suggest that radio-luminous AGNs 
play a more significant role in the cluster environment than they do in the field.  

If radio-excess is an indicator of AGN activity (see \S 6.3), then the excess observed in 
Figure~\ref{fig:lf14} is predominantly due to the early-type radio galaxies, which 
account for $\sim 70\%$ of the radio-excess objects (cf. \S 4.2).  To quantify this 
further, the radio LF for late-type galaxies only is computed and shown in 
Figure~\ref{fig:lf14}.  While early-type galaxies contribute to the luminosity density 
for the entire luminosity range considered here, the difference between the LF for the 
late-type galaxies and the LF for the aggregate sample increases for 
$\log L_{1.4\rm GHz} > 22.0$, suggesting an increasing contribution 
of radio emission from early-type galaxies at high luminosities.  The $3$ galaxies 
for which $\log L_{1.4\rm GHz} > 24.5$ are early-type Core galaxies.  
Despite the uncertainty involved in normalizing the cluster LF, no amount of 
reasonable normalization can account for the excess of radio sources appearing for 
$\log L_{1.4\rm GHz} > 24.5$---this would require a factor of at least $10^3$ difference
in the cluster and field LF normalizations, and the observed factor is $\sim 180$.  

\subsection{Faint Cluster Sources}

The disparity between the faint-end slope of the IR cluster galaxies and the YRC 
field galaxies may indicate that (1) we do not have a good constraint on the sampled volume for 
lower luminosity sources or (2) the cluster LF is truly flatter than the field LF
at low luminosity.  Explanation (1) can be illustrated by the observation that 
fainter sources in AWM7 are not detected as compared with A262 and A426, which just encompass 
AWM7 in redshift space (see Figure~\ref{fig:lumvdist}).  The statistics of AWM7 sources are
poor, and the sampled volume may be much smaller than the one taken here.  
Statement (2) might be true if the number counts of faint sources in clusters is lower 
than in the field due to the fainter and less massive sources accreting or merging with 
larger galaxies resulting in a top-heavy galaxy mass function.  

The IR LF alone indicates a flattening of the faint-end slope when compared with the field.
This flattening is unlikely due to NED selection since our
flux cutoffs are low enough to include normal field galaxies where optical and IR
selection are identical (e.g., Condon et al. 2002, YRC).  The effect responsible for
the flattening of the faint-end slope (i.e., selection effects, top-heavy galaxy mass function,
etc.) of the IR LF should also effect the radio LF faint-end slope in the same manner,
when we take into account the radio-luminous Virgo sources with $\log L_{60\mum}$ between 
$7$ and $8$.  We do not currently have the means to ascertain the reason for such an effect, 
but merely suggest it for future study.

\section{Discussion}

\subsection{Density-Morphology Relation}

The density distribution computed from deprojected late-type Ring galaxies shows no 
trend with radius and is scattered around a mean of $\sim 8$ Mpc$^{-3}$.  
The computed density of late-type galaxies within the Core is only $2.0$ Mpc$^{-3}$, 
$\sim 15$ times under-dense than the composite cluster, and a factor of $14$ smaller 
than the density of early-type galaxies in the Core region ($28.4$ Mpc$^{-3}$).  
The overall surface density of ellipticals in the Ring region is $2.86$ Mpc$^{-2}$, compared with 
a surface density of $19.10$ Mpc$^{-2}$ in the Core region.  These
observations indicate that early-type galaxies are more strongly clustered
in the Core than late-type galaxies.  The bias of early-type galaxies and the
analytic fit to the composite cluster density (\S 3) indicate that the high detection 
rate of ellipticals is not due to sample selection effects, but represents a
true increase in density.

The density-morphology relation is well established for optically-selected cluster
galaxies, but ours is a highly biased subset based on radio and FIR emission.
A valid point of inquiry therefore is whether the increase in early-type galaxy
density in the Core region is due to an underlying change in the density-morphology
relation for radio- and FIR-selected cluster galaxies, or if it simply reflects 
the existing density-morphology relation for all cluster galaxies and a random, 
uniform appearance of radio sources among them.  To answer this, we note that the 
exacting selection criteria used in this study makes it
likely that the radio and FIR ellipticals in our sample are a subset of the optically-identified
ellipticals.  The total fraction of detected early-types (cD, elliptical, and S0) in 
both the flux- and luminosity-limited samples is $\sim 30\%$, somewhat lower than the 
fraction typically found in ``spiral-rich'' clusters ($\sim 50\%$).  Consequently, the  
detection rate of early-type galaxies in our survey is consistent with a random population of 
radio sources tracing the underlying density-morphology relation for optical cluster galaxy 
samples.  The distribution of late-type galaxies can be understood in a similar manner to
that of the early-type galaxies.  Optically-selected samples show a late-type fraction that 
decreases toward the central regions of clusters.  The same correlation is found in our 
sample (see above).  Our flux cutoff is low enough to include galaxies with star formation 
rates similar to those found in the field, and the $60\mum$ selection is identical to the 
optical selection of late-type galaxies (Condon et al. 2002, YRC). 

\subsection{Mass Segregation in Clusters}

Mass segregation among cluster galaxies may induce the observed correlations between radio and FIR
luminosity and clustercentric radius as more massive galaxies may be more efficient IR and radio
emitters.  To investigate this possibility, 
$K$-band fluxes for all candidate sources were obtained from the Two Micron All Sky 
Survey (2MASS), using $K$-band luminosity as a proxy for mass.  The $K$-band luminosity to mass ratio 
is thought to be more universal than optical ratios, subject to lower 
extinction effects, and linked to older stars.  The 2MASS survey data include accurate magnitudes 
for sources less than $101''$ in extent.  

The cluster A1060, a few sources in AWM7 and Virgo, and almost all sources in Coma
were observed by 2MASS, but have either not been cataloged, or, as in the case
of Virgo, have not been mosaiced to obtain accurate $2.2\mu$m fluxes.  For
example, spatially integrated $2.2\mu$m fluxes are not reported for unmosaiced
Virgo cluster galaxies greater than $101''$ in extent.  These are some of the most
massive galaxies in our sample.  To assess possible mass segregation among Virgo
galaxies, we have used the data obtained by Boselli et al. (1997), which includes
$K'$ magnitudes measured from Infrared Space Observatory (ISO) images of $102$
Virgo cluster galaxies.  The data for Virgo are analyzed seperately from the 
remaining sample to preclude any systematic bias owing to differences in wavelength 
coverage and response between the 2MASS $K$-band and the ISO $K'$-band filters. 

\begin{figure}[hbt]
\centerline{\epsfxsize=8cm\plotone{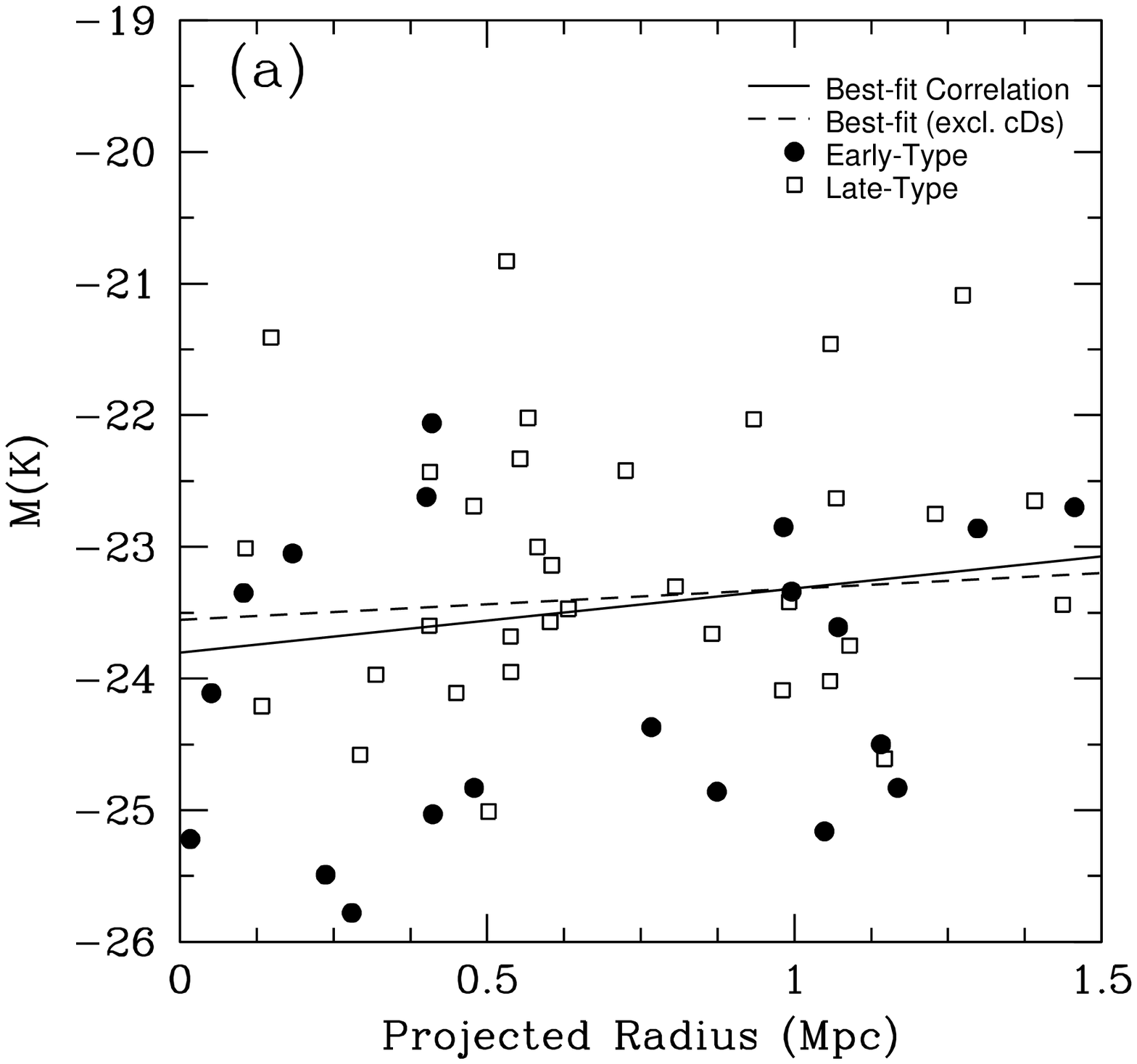}}
\centerline{\epsfxsize=8cm\plotone{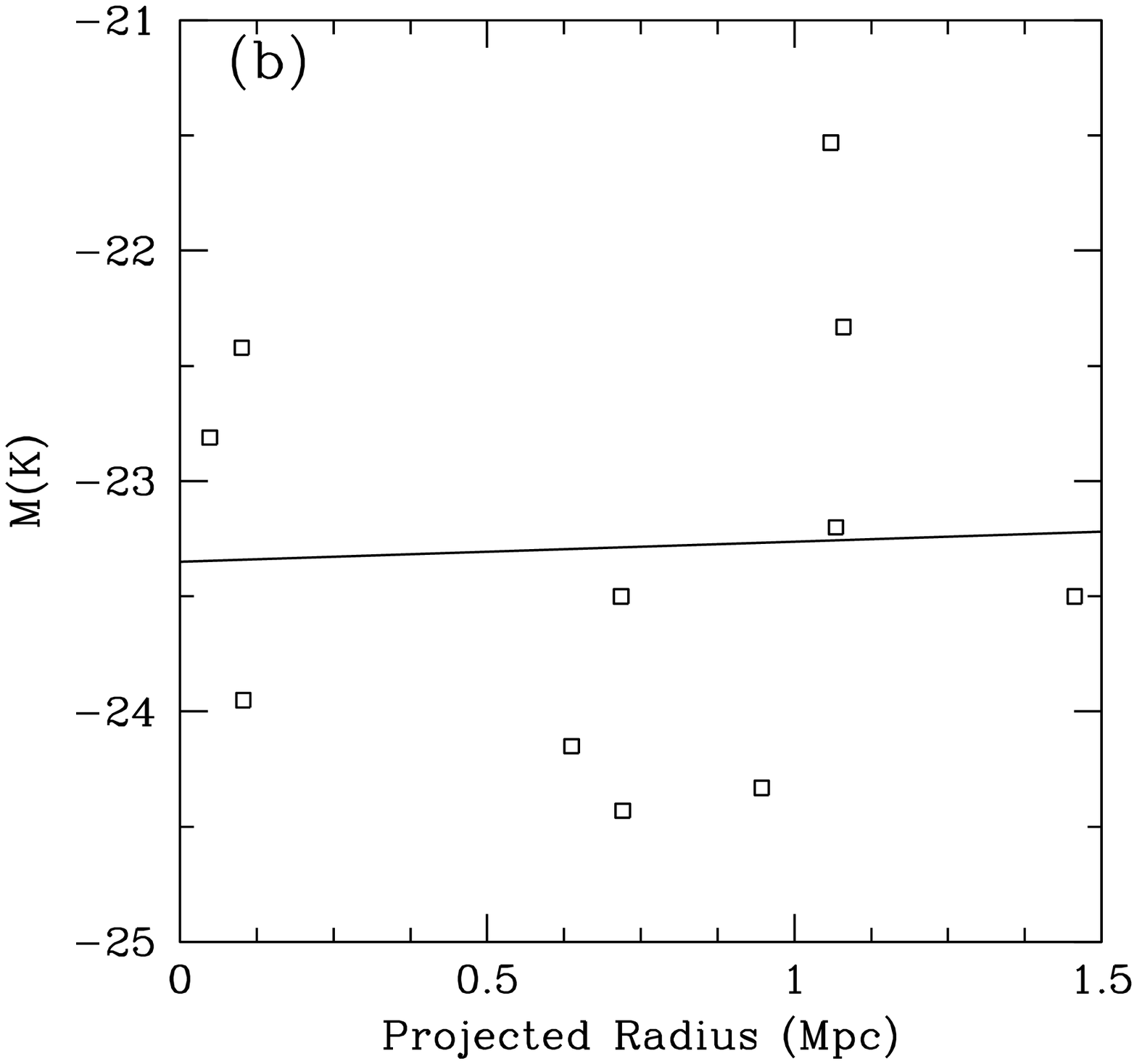}}
\caption{(a) Absolute $K$-band magnitude as a function of radius for the sample of $54$ galaxies with
reported 2MASS $K$-band fluxes.  Filled circles are early-type 
galaxies and open squares are late-type galaxies.  The solid and dotted lines indicate the 
best-fit correlations for the sample with and without cD galaxies, respectively.
(b) Same as (a), but for the $11$ Virgo cluster galaxies in the Boselli et al. (1997) near-IR
imaging survey.  In both samples, only a very weak positive correlation is found, indicating
that massive galaxies in the composite cluster do not preferentially populate cluster cores.  
\label{fig:kband}}
\end{figure}

For the $54$ sources with reported 2MASS $K$-band magnitudes, the $K$-band absolute luminosity has been 
computed and plotted against projected radial distance (Figure~\ref{fig:kband}a).  A
formal fit shows only a very weak correlation with $\sigma = 1.11$ dex:  
a $\sim 45\%$ probability of a positive correlation between M(K) and $r$.  Separate $\chi^2$ 
analyses for early- and late-type galaxies find the correlation is even weaker: the
corresponding errors in the slope of the correlations allow for constant M(K).  
The slight correlation for the composite cluster may be due to the central cD galaxies, as
well as the Messier and Seyfert galaxies in Virgo with an extent of $<101''$
in diameter.  The formal fit was redone excluding the two cD galaxies (NGC~708 and NGC~1275) 
for which $K$-band data were available.  Some galaxies with very small cluster offsets, 
small velocity differences, or unusually low $q$ value were left in the sample if their 
optical morphology indicated a non-cD galaxy.   The resulting correlation is even 
weaker (Figure~\ref{fig:kband}a), and the slope is statistically consistent with zero ($0.24\pm 0.39$ with 
$\sigma = 1.07$ dex).  Eleven galaxies in our Virgo sample were imaged in the Boselli et al. (1997) 
study, including half of our detected Messier sources.  Their absolute $K'$-band magnitudes 
are plotted in Figure~\ref{fig:kband}b as a function of projected radial distance.  
As with the previous analysis, we find only a $\sim 45\%$ probability of a positive correlation 
between M(K) and $r$, and the statistics are consistent with a flat slope.

Luminosity segregation in clusters has been studied extensively, leading to
mass estimates by invoking some M/L ratio.  The mass 
per galaxy can be obtained by normalizing the luminosity density by the number density.
As an example, we consider the Virgo cluster.  The luminosity density profile has an identical
slope to the number density distribution in Virgo, except in the very central ($r<60'$)
region of the cluster, where the luminosity density is steeper than the
number density (Schindler et al. 1999).  Neglecting these central bright galaxies 
(most of which are Messier and Seyfert galaxies), the inferred 
average mass per galaxy is constant with radius.  Again, this conclusion 
assumes some constant M/L ratio based on the average M/L ratio of early- and late-type
galaxies in clusters.  As another example, there is marginal evidence for segregation 
in the inner $20\%$ of the core radius of Coma, but no large-scale correlation is found 
(Zhao et al. 1991; Bahcall 1973 and references therein).  Inclusion of Coma galaxies 
in Figure~\ref{fig:kband} should therefore not significantly affect our conclusion based 
on the clusters for which $K$-band data were available.  The absence of mass
segregation for the time-averaged evolution of the composite cluster is important for our study 
since it indicates that radial variations in radio and FIR emission are not a result of gas-poor 
massive ellipticals dynamically sinking into the Core region.

\subsection{Discerning AGN from the Normal Population}

Determining the power source of cluster galaxies is crucial in order to correctly 
interpret the radio and FIR correlations discussed in previous sections.  Our sample is
small enough that a case-by-case analysis of the multiwavelength data for each galaxy can
be carried out to determine which show evidence of AGN.  We have
searched the literature for each of the radio-excess sources, and we assign an AGN
classification based on evidence of hard X-ray emission, radio jets, or broad optical
emission lines consistent with an active nucleus.  This resulted in $5$ AGN
classifications.  Miller \& Owen (2001a) obtained high-resolution data for
two of our $4$ remaining radio-excess sources (CGCG097-073 and KUG1258+287) and
were classified as star forming galaxies based on their nuclear optical spectra.

The validity of using radio-excess as an indicator of AGN has been explored in previous
studies (e.g., YRC; Yun et al. 2000; Condon \& Broderick 1988).  
The exact AGN fraction among radio-excess sources in the Miller \& Owen (2001a) sample
is unclear as (1) they were unable to obtain spectroscopy for all of their sources, 
(2) their spectroscopy cannot distinguish a weak AGN system from a compact
starburst dominated emission-line spectrum, and (3) they include galaxies in the AGN sample 
which show evidence of an old stellar population from the absorption-line spectrum 
(Miller \& Owen 2001a).  They test the effects of (2) by obtaining high-resolution ($\sim
0.25''$) radio images of cluster galaxies in A1367 and A1656, and find that of the $5$
of $20$ detected galaxies, at least $4$ have a compact nucleus.  Of the remaining
$15$, $13$ are known star forming galaxies whose nebular emission may overwhelm the emission 
associated with an AGN.  

We have independently analyzed the data of Miller \& Owen (2001a) and find that
of the $\sim 70$ spectroscopically-confirmed star forming galaxies in their 
radio-selected sample, only 
$6$ show radio-excess, whereas $\sim 23/77 = 30\%$ of their 
spectroscopically-confirmed AGNs show radio excess.  The overall frequency of AGN sources
among their radio-excess galaxies is therefore $\sim 80\%$.  It is likely that
the actual frequency of AGNs among the radio-excess galaxies is higher, since few optical
spectra were collected for the most radio-luminous sources ($q<1.00$).
Miller \& Owen (2001b) detect 4 FIR luminous ($\log L_{\rm FIR} \geq 11.0$) galaxies.  
Of these, one is an IR-excess galaxy with $q=3.21$ and the rest have nominal $q$ values.
None of the IR-excess sources in our sample show evidence for AGN activity. 
The $13$ radio-excess sources in our sample include 4 AGN cD galaxies and 5 non-cD AGN sources.  Hence,
of the radio-excess sources in our sample, $\sim 70\%$ are 
AGN based on the presence of radio jets, X-ray emission, and/or optical emission lines.
This fraction is a lower limit since some of the less well-studied radio-excess objects could
also host an AGN.  In summary, it appears that a significant fraction of radio-excess objects
are associated with luminous AGN.

\begin{figure}[hbt]
\plotone{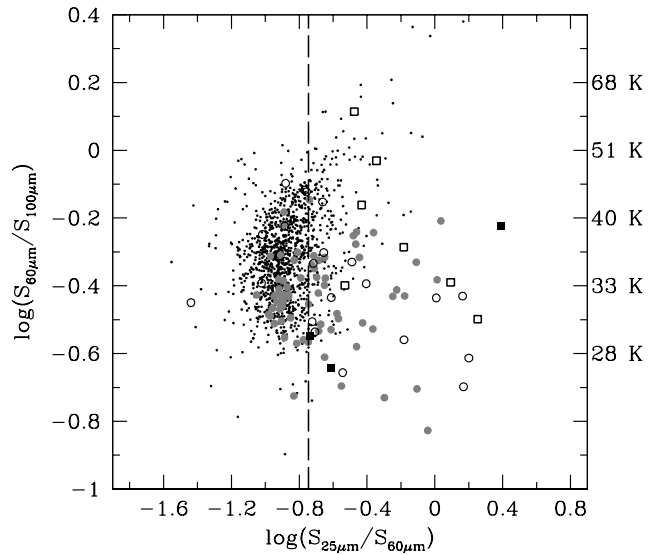}
\caption{IRAS FIR versus mid-IR color-color diagram for those sources in the
luminosity-limited sample with available
IRAS $25\mu$m measurements.  FIR color has been converted to temperature using
$\beta = 1$.  The large open and filled grey circles represent early- and 
late-type galaxies, respectively.  The radio-excess sources are shown with large
open and filled squares for early- and late-type galaxies, respectively.  The small
points are for the YRC field galaxies with distance less than $94$ Mpc.  The hashed
vertical line denotes the proposed line indicating the lower boundary for IR 
Seyfert galaxies (de Grijp et al. 1985), and radio-excess sources are found to the
right of this line.  This suggests that the mid-IR excess which characterizes these
sources is attributable to dust heating in the circumnuclear environment surrounding
an AGN.
\label{fig:color}}
\end{figure}

A segregation between IR luminous, normal, and  radio-excess objects can be seen by
examining the IRAS FIR--vs.--mid-IR color-color diagram (see YRC for a discussion).
The comparison of the environment in radio-excess cluster objects to ``normal'' 
cluster galaxies can be investigated in the same way.
In Figure~\ref{fig:color}, the large open and filled grey circles represent early- and 
late-type cluster galaxies with $q>1.64$, respectively.  The radio-excess sources
are shown with large open and filled squares for late- and early-type galaxies,
respectively.  For comparison, the distribution for the IR-selected YRC field
galaxies within $94$ Mpc is indicated by the small black points.

The aggregate sample of cluster galaxies in the luminosity-limited sample have 
$\langle S_{60\mum}/S_{100\mum}\rangle \sim 0.42 \pm 0.17$, and the appropriately selected 
field galaxies have $\langle S_{60\mum}/S_{100\mum}\rangle \sim 0.49 \pm 0.07$.  For 
comparison, the mean FIR color in the cluster sample of Bicay \&
Giovanelli (1987) is $\langle S_{60\mum}/S_{100\mum}\rangle \sim 0.38$, slightly lower, but
still consistent with our result.  The slight difference may be due to our selection of
cluster galaxies based on FIR emission, biasing the sample to galaxies with warmer FIR 
color than seen in optically-selected galaxies.  

All of the radio-excess sources which have detected $25\mu$m emission
appear to the right of the proposed line indicating the lower boundary for IR Seyfert galaxies
according to their $S_{25\mum}/S_{60\mum}$ colors (de Grijp et al. 1985).  
Sources detected at $25\mu$m will have higher $S_{25\mum}/S_{60\mum}$ ratios in general than those
not detected at $25\mu$m.  The absence of radio-excess sources to the left
of the proposed indicator for IR Seyfert galaxies is not significant by itself.  However, when
taken in conjunction with the observation that most radio-excess galaxies in the field have
warmer mid-IR colors than the normal population by $\sim 0.5$ dex (YRC), this suggests
that radio-excess sources in the cluster also, on average, have warmer mid-IR colors, 
consistent with a model in which dust is heated in the circumnuclear region by an AGN.
The early-type radio-excess objects (most of which are AGN)
populate the same region in color-color space as the radio-excess field objects 
appearing to the right of the IR Seyfert line (YRC). 
In summary, radio-excess sources show mid-IR colors consistent with AGN, and this
is supported by the case-by-case analysis.  The segregation of $q$ values indicates that
most of these radio-excess sources are in the Core region.  

While radio-excess sources generally show mid-IR excesses, the opposite is not true.
Namely, there are many late-type galaxies that do not have a radio-excess but which do 
show mid-IR-excess.  This is not surprising since either a young or compact starburst or 
radio-quiet AGN could result in the observed mid-IR color (e.g., Efstathiou, Rowan-Robinson, 
\& Siebenmorgen 2000).  We can place an upper
limit to the fraction of mid-IR excess cluster sources with $q>1.64$ of $51\%$, with the
field value being $<10\%$.  A direct comparison of the two fractions is risky as there may
be a number of selection effects we have not considered.  If, in fact, a larger fraction of 
normal cluster galaxies exhibit mid-IR excess when compared with the field, it may indicate
that many of the disk galaxies host young, dust-enshrouded starbursts.  If so, this provides
some marginal evidence for harassment-induced star formation activity (see \S 6.6).

\subsection{FIR Emission Mechanisms}

\subsubsection{X-Ray Dust Heating}

\begin{figure}[hbt]
\plotone{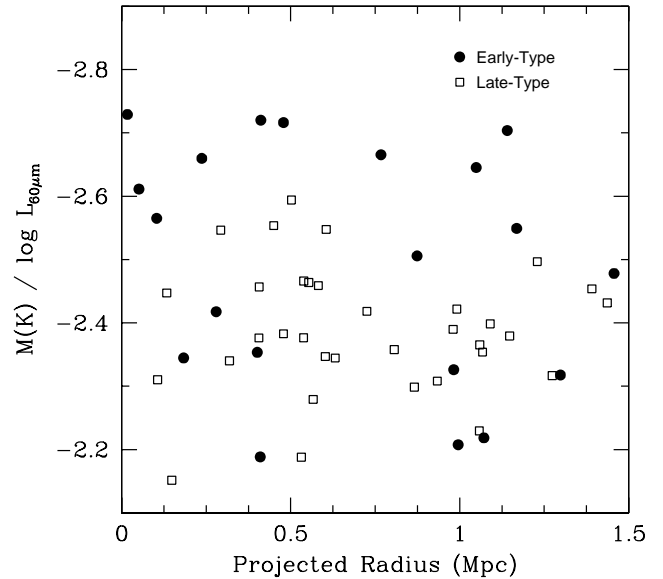}
\caption{Ratio between $K$-band absolute magnitude and log of $60\mu$m luminosity
versus clustercentric distance for the luminosity-limited sample. 
Solid circles are early-type galaxies and
open squares are late-type galaxies.  We find no significant correlation
between the two, indicating that electron-dust collisions have a relatively
insignificant effect on the dust heating rate for most galaxies 
in the cluster environment (see text).
\label{fig:dustheat}}
\end{figure}

According to \S 6.2, the composite cluster shows no evidence of mass segregation based
on the galaxies' $K$-band luminosity.  If only the cirrus component of dust is stripped from
a galaxy, then this will have an effect on the observed emission at $100\mum$, but the net
impact on $L_{\rm FIR}$ should be small.  This argument is difficult to quantify since it
requires a knowledge of the dust distribution within galaxies.  However, dust is closely 
associated with molecular gas which can be accurately traced by millimeter CO transitions.
Young et al. (1995) have shown that more than $90\%$ of the CO emission from late-type galaxies
comes from the inner $2/3$ of the optical disk, and the same may also be true of FIR emission.  
Consequently, the mass-to-light ratio as traced by M(K)/L$_{60\mum}$ should be insensitive
to the effects of dust stripping.  The presence of significant collisional
heating should result in the more centrally located galaxies having lower M(K)/L$_{60\mum}$ 
values.  Figure~\ref{fig:dustheat} shows no correlation between M(K)/L$_{60\mum}$ and radius
indicating that collisional heating due to electrons is not a significant source of
FIR emission.

\begin{figure}[hbt]
\centerline{\epsfxsize=8cm\plotone{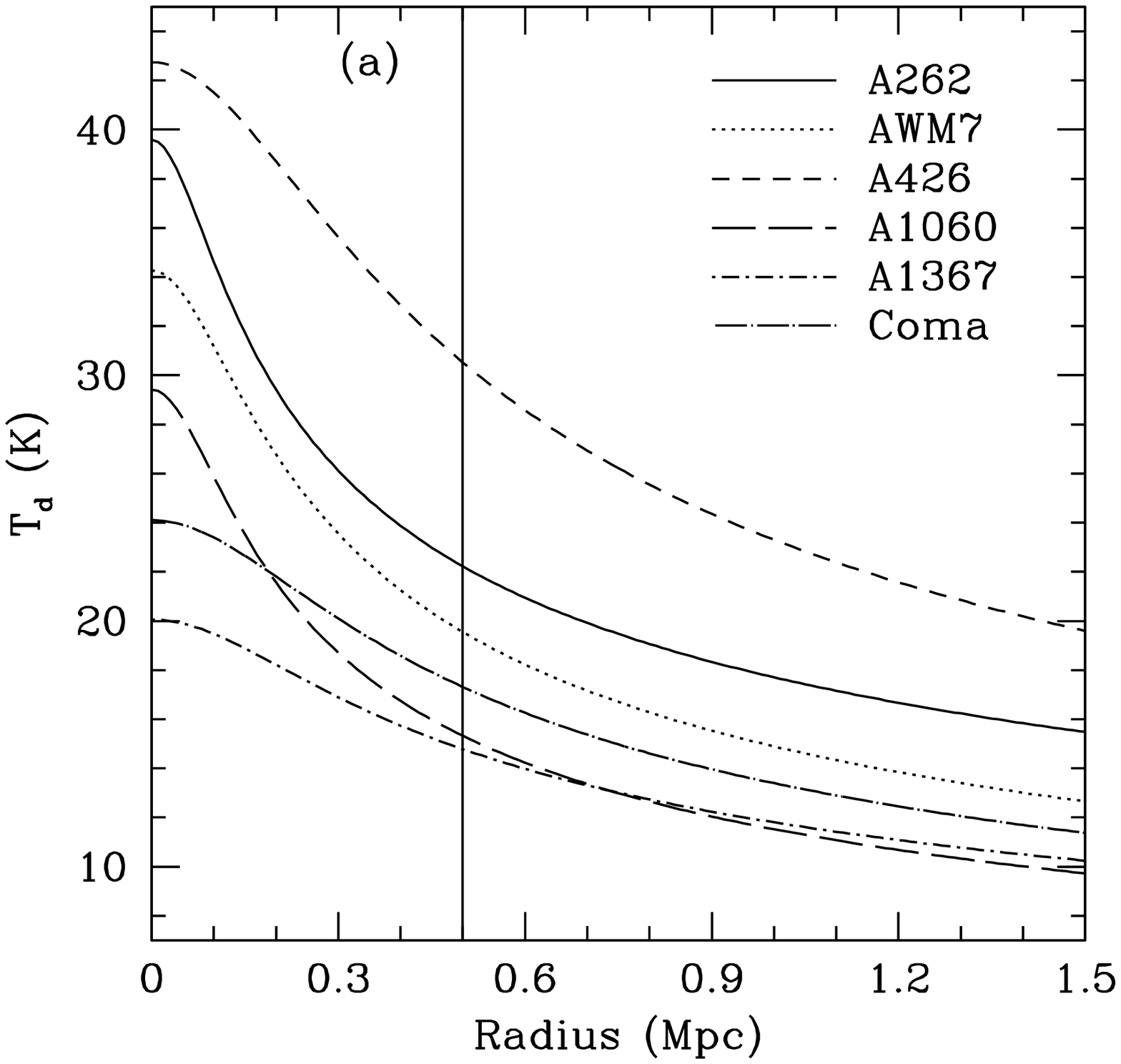}}
\centerline{\epsfxsize=8cm\plotone{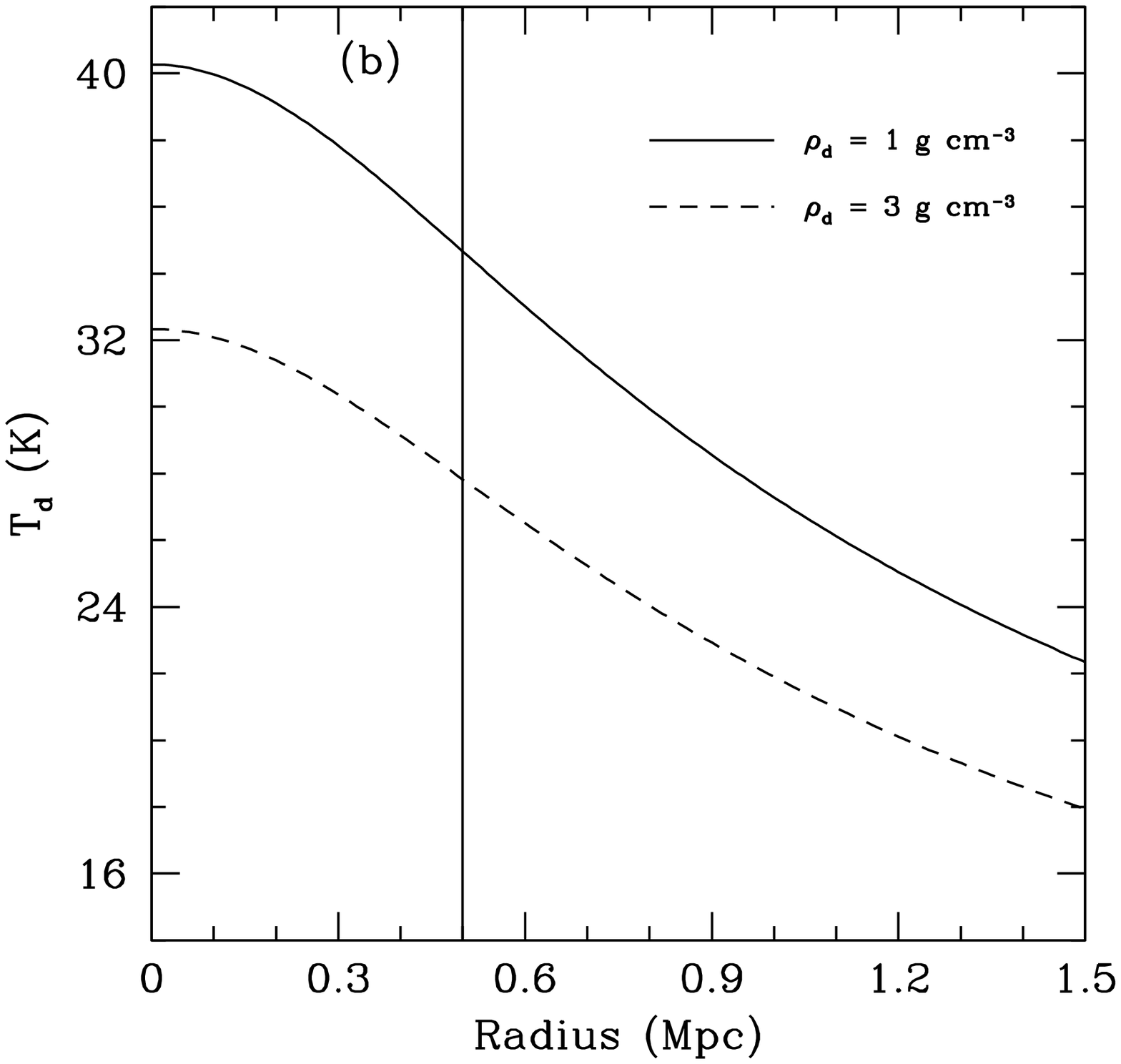}}
\caption{(a) Dust temperature versus radius for a collisionally-heated grain population 
of size $a=10^{-5}$ cm and grain density $\rho_{\rm d} = 3$ g~cm$^{-3}$ for 6 of the
7 sample clusters, using the $\beta$-model parameters derived from the X-ray surface
brightness profile (Mohr, Mathiesen, \& Evrard 1999).  (b) 
Dust temperature versus radius for a collisionally-heated grain population
of size $a=10^{-5}$ cm for the composite cluster, adopting the $\beta$-model parameters
from \S 3.  The solid and hashed curves correspond to $\rho_{\rm d} = 1$ and $3$ 
g~cm$^{-3}$, respectively.  The solid vertical lines in Figure~\ref{fig:xrayprof}a
and Figure~\ref{fig:xrayprof}b denote the boundary between the Core and Ring
regions.\label{fig:xrayprof}}
\end{figure}

The effects of collisional heating of dust by ICM electrons
can be estimated by calculating the collisional heating rate.  Following the analysis
of Goudfrooij \& de Jong (1995), the heating rate is given by
$$H_{\rm e} (r) = n_{\rm e}(r)\left(\frac{8kT_{\rm e}}{\pi m_{\rm e}}\right)^{1/2} 
\left(\frac{3}{2} kT_{\rm e}\right)\pi a^2,\eqno(9)$$
where $T_{\rm e}$ is the electron temperature, $a$ is the dust grain radius, and 
$n_{\rm e}(r)$ is the radial electron density profile, assumed to follow a $\beta$ profile
similar to equation (5).  The dust temperature is obtained by equating the collisional
heating rate to the cooling rate due to IR emission given by de Jong et al. (1990), yielding
$$T_{\rm d} = \left(\frac{n_{\rm e}(r)(8kT_{\rm e}/\pi m_{\rm e})^{1/2}
(3kT_{\rm e}/2)}{6.0\times 10^{-8} \kappa_{\rm o} (4a \rho_{\rm d}/3)}\right)^{1/5},\eqno(10)$$
where the opacity $\kappa_{\rm o} = 2.4\times 10^3$ (Hildebrand 1983), $\rho_{\rm d}$ is
the grain density, and a dust emissivity proportional to $\lambda^{-1}$ has been assumed.
Assuming a typical value of $T_{\rm e} \sim 10^7$ K, the temperature-radius curves for
6 of the 7 clusters (excluding Virgo) are shown in Figure~\ref{fig:xrayprof}a, for the values
of $r_{\rm c}$, $\beta$, and $n_{\rm e}(0)$ derived by Mohr, Mathiesen, \& Evrard (1999)
from the X-ray surface brightness profiles.  The curves shown are those corresponding to
$a = 10^{-5}$ cm and $\rho_{\rm d} = 3$ g~cm$^{-3}$.  The temperature-radius relation is
sensitive to the product of $a$ and $\rho_{\rm d}$, and Figure~\ref{fig:xrayprof}b shows
the curves corresponding to different grain densities assuming the $\beta$-model parameters
derived in \S 3 for the composite cluster.  

The effects of X-ray dust heating do appear to be significant in cD galaxies containing 
large amounts of hot X-ray gas associated with the ICM .  
In this case, the electron density is high enough 
that dust heating becomes important (e.g., Bregman, McNamara, \& O'Connell 1990; 
Lester et al. 1995) and the emission is uncharacteristic of the localized 
FIR emission associated with active star formation.  This can be seen in 
Figures~\ref{fig:xrayprof}a and \ref{fig:xrayprof}b, where the dust temperature at the center 
of the cluster is $\sim 40$K, $21\%$ higher than the average temperature derived 
for cluster galaxies using the FIR color alone.  

Figure~\ref{fig:dustheat} indicates that collisional heating of dust grains
in the sample galaxies has a negligible effect on the observed FIR emission, allowing us to
place a loose constraint on the dust column density.  For example, the mean FIR color temperature
for the cluster galaxies is $\sim 33$K (see Figure~\ref{fig:color}), indicating 
$a\rho_{\rm d} > 2.7\times 10^{-5}$ g~cm$^{-2}$. For the typical values of $a$ and 
$\rho_{\rm d}$ in Figure~\ref{fig:xrayprof}b, the derived dust temperatures are not 
sufficient to account for the observed FIR color temperature, particularly in the Ring
region, where $T_{\rm d}$ falls below $30$ K.  If we assume a limiting value of
$a\rho_{\rm d} = 2.7\times 10^{-5}$ g~cm$^{-2}$, then the collisional heating rate 
is $\lesssim 50\%$ of the optical heating rate at $r=0.5$ Mpc and $\lesssim 6\%$ of 
the optical heating rate at $r=1.5$ Mpc.  The lower limit of $a\rho_{\rm d}$ given 
above indicates that large dust grains are primarily responsible for the FIR emissivity.

\subsubsection{Gas and Dust Stripping}

Spiral galaxies in rich clusters show a marked HI gas deficiency when compared 
with isolated field spiral galaxies (Giovanelli \& Haynes 1985; Cayette et al. 1990).
The mechanisms used to explain stripping include ram pressure (Gunn \& Gott 1972),
turbulent viscosity (Nulsen 1982), tides during galaxy-galaxy interaction
(e.g., Richstone 1975; Merrit 1983), or thermal evaporation (Cowie \& Songaila 1977).
More recent studies have suggested that the inner disk of molecular gas remains intact
as it is more strongly bound to the galaxy 
(Kenney \& Young 1986; Cayatte et al. 1994; Perea et al. 1997; Bicay \& Giovanelli 1987).
Contursi et al. (2001) have found that HI deficient galaxies in Coma and A1367 generally have
comparable dust masses to HI-normal galaxies.  

The FIR emission for cluster galaxies in our sample is independent of radial distance 
in agreement with that found in other cluster samples (e.g., Rengarajan, Karnik, \& Iyengar 1997).  
The apparent flatness of the FIR luminosity-radius
correlation does not address directly the question of the gas and dust content of
the galaxies since a system with globally low gas content could have 
localized star formation producing the same levels of FIR emission as a gas-rich 
system with modest levels of global star formation.  While higher resolution
data can better address this question, optical (e.g, H$\alpha$), mm (e.g.,
CO), and radio (e.g., HI 21 cm) tracers indicate that galaxies have 
lower cold gas content toward the central regions of clusters.  This trend, if
it is exists in our sample, does not affect the observed FIR emission.

\begin{figure*}
\centerline{\epsscale{1.6}\plotone{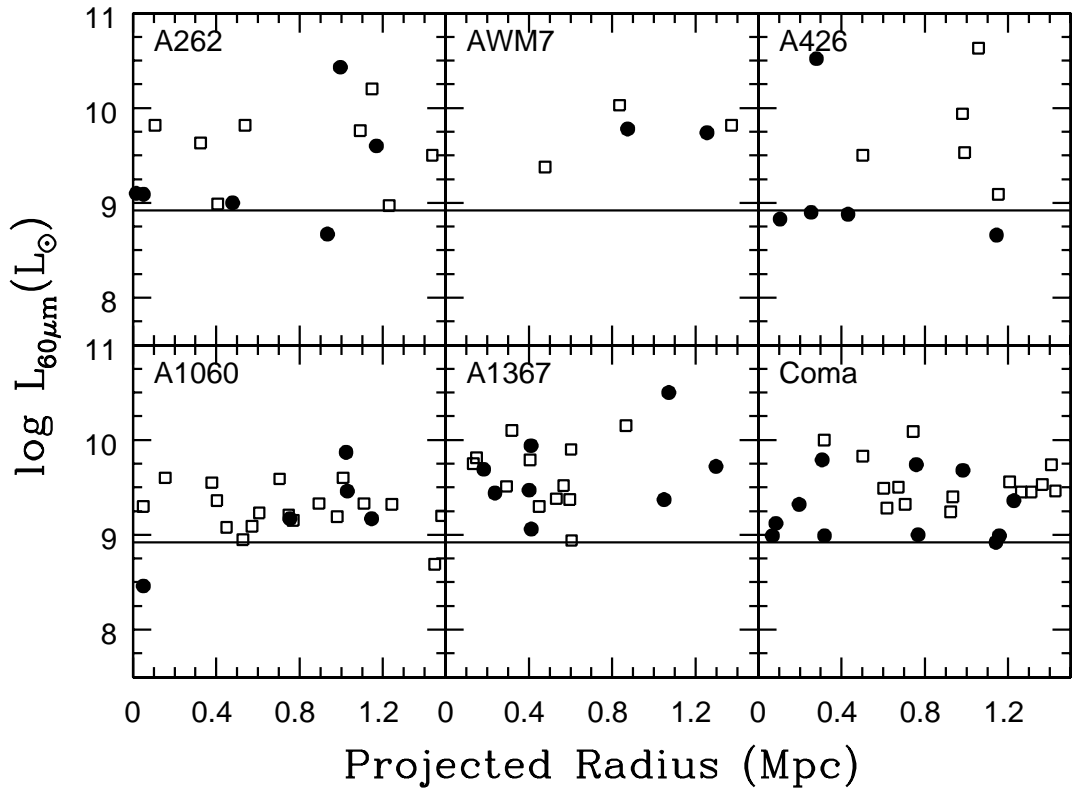}}
\caption{IR luminosity versus projected radius for $6$ of the $7$ clusters in the
flux-limited sample.  The 
horizontal line in each plot denotes the completeness limit.  Filled circles and 
open squares represent early- and late-type galaxies, respectively.  We find no
significant cluster-to-cluster variation in the IR luminosity as a function of radius.
\label{fig:radial}}
\end{figure*}

The flat FIR luminosity-radius correlation 
seen in Figure~\ref{fig:lumvrad} suggests that changes in the environment within $1.5$ Mpc
do not have a significant impact on the FIR emission.   The cluster-to-cluster variation 
in the FIR luminosity--radius correlation is shown in Figure~\ref{fig:radial} for $6$ of the
$7$ clusters.  A262 and AWM7 have few detected sources above our luminosity cutoff, and any trend 
will be unclear given the few data points for these two clusters.  For the remaining clusters, the FIR 
luminosity is relatively flat with radius.  

\begin{figure}[hbt]
\epsscale{1.0}
\centerline{\epsfxsize=8cm\plotone{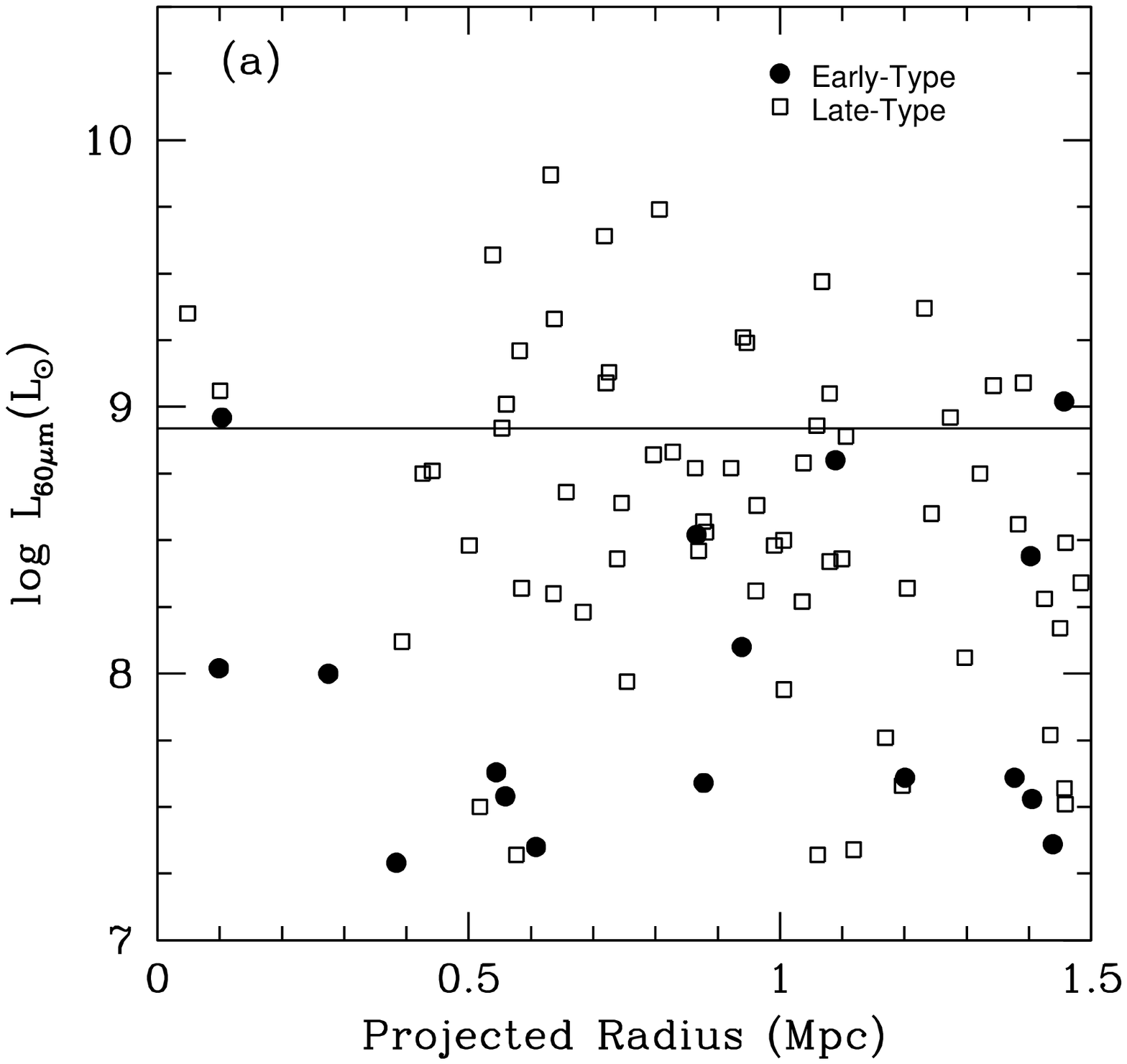}}
\centerline{\epsfxsize=8cm\plotone{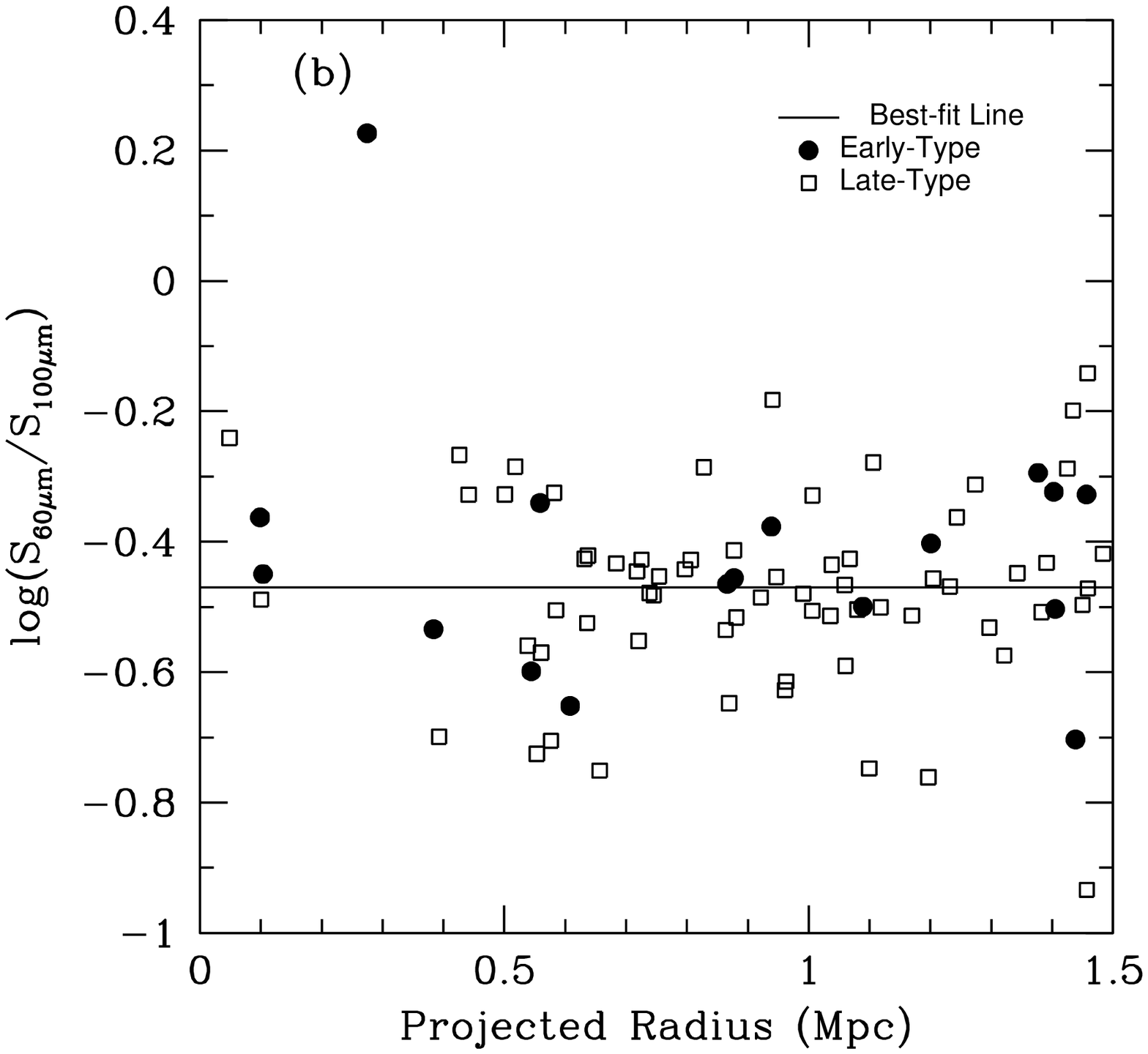}}
\caption{(a) IR luminosity versus projected radius for the flux-limited sample of Virgo galaxies.  Filled circles
and open squares indicate early- and late-type galaxies, respectively, and the solid horizontal
line denotes our luminosity limit.  (b) IRAS FIR color as a function of projected radius for the flux-limited 
sample of Virgo
galaxies.  Filled circles and open squares indicate
early- and late-type galaxies respectively.  There is no correlation between FIR color and
radius for both early- and late-type galaxies.  Also note the paucity of IR sources in Virgo
for radii $<0.5$~Mpc from the cluster center.\label{fig:virgofir}}
\end{figure}

The distinct lack of detected radio and FIR sources in the Virgo cluster
core could be a result of severe FIR suppression (Figure~\ref{fig:virgofir}a).  This reduction 
in FIR luminosity is correlated with HI deficiency in the Virgo galaxies studied by
Doyon \& Joseph (1989)---many of which fall within our Core region---and who
quote a factor of 3 decrease in star formation as a result.  The recent H$\alpha$ imaging
study of Virgo spiral galaxies by Koopmann \& Kenney (2002) indicates that almost all
show evidence for truncated star formation.  The FIR emission for Ring Virgo galaxies
appears to weakly decrease with radius (Figure~\ref{fig:virgofir}a), and may be expected if the 
diffuse cold dust component is stripped in the higher
density regions of the cluster, leaving only the warm component to dominate
the FIR luminosity.  To test for this effect, we compute $S_{60\mum}/S_{100\mum}$ (FIR color)
versus radius for the flux-limited sample of Virgo late-type galaxies (Figure~\ref{fig:virgofir}b).
Stripping of cold dust should yield warmer FIR colors toward the cluster center, but this is
not seen in Figure~\ref{fig:virgofir}b.  We conclude that cold dust stripping, if it exists
in the Virgo cluster, does not affect the FIR emission.

\subsection{Radio Enhancement}

The $q$ value is not affected by any increase or decrease in the star formation rate of a 
galaxy as the radio and FIR emission are directly proportional to this rate.  Our observation 
of a $q$-radius correlation and lack of FIR luminosity-radius correlation (\S 4.3) indicate that 
changes in star formation rates are not responsible for the increase
in radio luminosity of sources toward the center of the composite cluster.  
Tidally-induced star formation is therefore not a dominant
mechanism in increasing the radio emissivity of sources toward the center clusters.

While there are relatively few mechanisms that can only affect the observed FIR emission,
radio enhancement with nominal FIR emissivity can occur in a number of ways,
including ram-pressure and thermal magnetospheric compression, and elevated AGN
activity.  Each of these possibilities is discussed below.

\subsubsection{Ram Pressure Magnetospheric Compression}

Ram pressure magnetospheric compression can occur if a galaxy has sufficient velocity
through the ICM causing the gas on the leading edge of the galaxy to be compressed.
The magnetic field coupled to the gas will also be compressed, resulting in increased
synchrotron emission.  Alternatively, thermal compression
occurs when the pressure of the ICM itself is enough to compress the magnetic field to
the point where it may affect the observed synchrotron emission (see below).
 
The effects of ram pressure are assessed by computing
the line-of-sight velocity deviation of each galaxy from the cluster 
velocity normalized by the cluster velocity dispersion:
$$\delta v = {|v_{galaxy} - v_{cluster}| \over \sigma_{cluster}},\eqno(11)$$
referred to as the ``relative velocity'' of each galaxy (see Miller \& Owen [2001a], eq. [3]).
We have obtained the cluster velocity dispersion data for the Abell sources 
from Struble \& Rood (1999).  The velocity dispersion of AWM7 and Virgo
were obtained from Koranyi et al. (1998) and  Fadda et al. (1996), respectively.  
The velocity dispersion of AWM7 shows a weak radial profile, with 
${\rm d}\sigma \sim 150$ km~s$^{-1}$ lower in the inner $0.13$ Mpc.  Since our sample 
extends to $1.5$ Mpc, we adopt the dispersion velocity found for the 
outskirts of the cluster ($\sigma \sim 680$ km~s$^{-1}$).  A summary of the
adopted cluster velocity dispersions is given in Table~\ref{tab:clsum}.

\begin{figure}[hbt]
\plotone{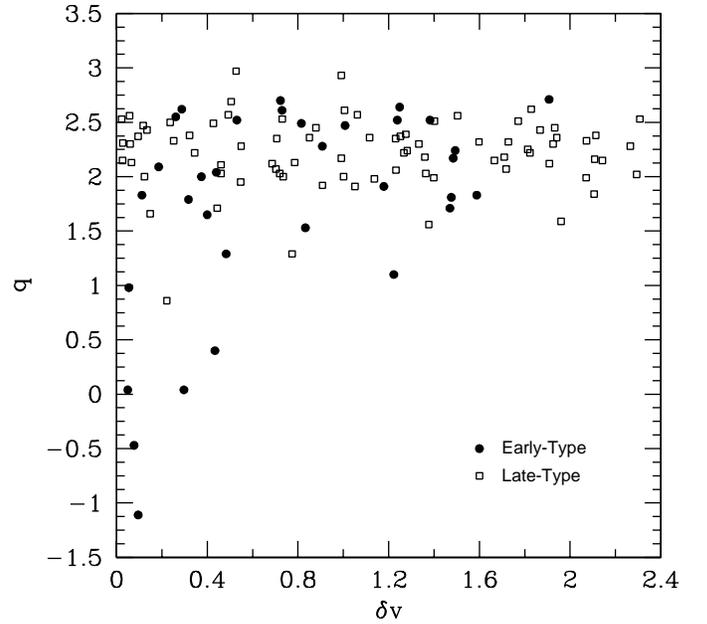}
\caption{$q$ parameter as a function of relative (projected) velocity, as computed from
equation (11), for the luminosity-limited sample.  Filled circles
and open squares represent early- and late-type galaxies, respectively.  We find only a 
very weak correlation between $q$ and $\delta v$.  The correlation, if it exists, may 
be masked by a number of effects (see text).
\label{fig:relvel}}
\end{figure}

Figure~\ref{fig:relvel} shows the distribution of $q$ values as a function of relative 
velocity for the cluster galaxies, with separate emphasis on
early and late-type galaxies.  The deficit of galaxies in the lower right-hand corner
of Figure~\ref{fig:relvel} is due to projection effects, since some of those galaxies
with a large cluster velocity may have small line-of-sight velocity if the direction
of motion is aligned perpendicular to the line-of-sight.

A significant contribution of ram pressure magnetospheric compression should manifest itself
as an anti-correlation between the $q$ distribution and relative velocity.  For early-type
galaxies, we find a slope of $0.56\pm 0.28$ (including radio AGNs).  Excluding all early-type 
galaxies with radio-excess, the slope becomes zero.  One might argue that excluding all radio-excess
sources is not appropriate since some of the radio-excess in these galaxies may be due 
to ram pressure magnetospheric compression, particularly in light of the efficiency for
ram pressure to increase synchrotron emission due to the $v^{2}$ dependence of the
magnetic field.  However, most of these sources {\it do} show evidence of powerful
AGN activity and most are massive ellipticals with rather low relative velocity 
(Figure~\ref{fig:relvel}).  Projection effects may also produce this effect, 
but it is unlikely that {\it all} radio-excess sources have velocities 
perpendicular to the line-of-sight.  Therefore, the removal of these radio-excess 
sources is warranted given the apparently distinct source for radio-emission.
Hence, there is no obvious trend indicating that increased velocity 
among elliptical galaxies results in increased radio emission.  

One caveat is as follows.  Low velocity ellipticals in 
dense environments will have a longer interaction time with neighboring galaxies.  The 
result may be increased gas accumulation in the major component of the interaction, 
leading to an increase in the intensity and/or duration of star formation 
activity and/or result in elevated AGN activity.  Our data suggest that tidally-induced star 
formation is not a major contributor to the enhancement of radio emission in most
non-cD elliptical galaxies.  Assuming this to be the case, it is unlikely that a longer 
interaction time is important in increasing $q$ values for low-velocity galaxies in the
Core region.  The low-density environment characteristic of late-type galaxies (cf. \S 3) in 
the Ring argues against the possibility of increased radio emission due to longer interaction time:
the velocity dispersion generally flattens out for large radii, whereas
the density falls off rather rapidly as a power law (cf. eq. [5]).

The late-type galaxies exhibit a weak correlation between $q$ and $\delta v$, similar 
to the findings of Miller \& Owen (2001a).  Other studies have indicated little correlation
between projected velocity and projected radius.  As an example, there is great 
scatter between velocity and radius in the Coma cluster with no obvious relation between the two 
(Biviano et al. 1995).  The statistics of our sample are such that we cannot make a secure determination 
for individual clusters, and the lack of a strong $q$-$\delta v$ correlation does not rule out the 
possibility that some of the radio emission from the cluster galaxies is due to ram-pressure enhancement 
of the magnetic field.  

There are clear cases where ram pressure enhancement has a noticeable effect on the synchrotron power 
(e.g., Cayette et al. 1990).  However, the effect, even if present on a large scale, may be difficult 
to observe.  In particular, Miller \& Owen (2001a) suggest several possibilities for non-gaussian 
motions in clusters, including bulk flows in clusters and cluster cooling flows.  Projection effects 
may also weaken such a correlation if it exists.  These effects, compounded with a lack of knowledge 
of the spatial coordinates of a galaxy with respect to the core of the cluster, all serve to muddle any 
correlation between $q$ and $\delta v$.   

\subsubsection{Thermal Compression}

The observed radio enhancement among star forming late-type galaxies could result from 
magnetic field compression due to the thermal pressure of the ICM.  Miller \& Owen (2001a) find that 
thermal pressure can account for most, if not all, of the pressure required to produce the observed 
radio enhancement in their sample.  The effects of thermal compression can be quantified by computing
the increasing in synchrotron power due to the compression of the magnetic field as a result
of external pressure.  Following the analysis of Miller \& Owen (2001a), the gas density distribution 
will follow the $\beta$ profile of equation (5), and the thermal pressure is 
$$P_{\rm th} = \frac{\rho kT}{\mu m_{\rm H}},\eqno(12)$$
where $T$ is the temperature of the gas and $\mu = 1$ for a gas of pure hydrogen.  The synchrotron
power is directly proportional to the pressure on the gas which, in the cluster, can be thought of
as the sum of two pressures:
$$P_{\rm tot} = P(B_o) + P_{\rm th} = \frac{B^2}{8\pi}.\eqno(13)$$
Here, $P(B_o)$ refers to the pressure due to the ambient magnetic field of the galaxy if it were not
located in the cluster, $P_{\rm th}$ refers to the thermal pressure provided by the ICM, and $B$ refers
to the value of the magnetic field that would produce such a total pressure.  The ratio of the 
synchrotron luminosity in the cluster ($L_{\rm c}$) to that of the same galaxy in the field ($L_{\rm f}$)
can then be written as
$$\frac{L_{\rm c}}{L_{\rm f}} = \frac{P(B_o)+P_{\rm th}}{P(B_o)} = 1 + 
\frac{\rho kT}{\mu m_{\rm H} P(B_o)}.\eqno(14)$$
The dashed curve in Figure~\ref{fig:lumvrad}b represents the expected value of $L_{\rm c}$ as a function of
radius, where we have made the following assumptions.  First, similar to Miller \& Owen (2001a), we have 
assumed the $\beta$-model parameters for Coma as deduced by Mohr et al. (1999), as it is most representative 
of relaxed clusters ($r_c = 0.257$ Mpc, $\beta = 0.705$, $\rho_c = 
7.4\times 10^{-27}$ g~cm$^{-3}$, and $kT = 8.21$ keV).  Secondly, we have assumed that a typical field 
spiral galaxy has a magnetic field of $\sim 5\mu$G. For these values, the thermal pressure is  
$\sim 1.35\times 10^{-12}$ dyn~cm$^{-2}$ at $r=1.5$ Mpc from the cluster center, approximately   
$1.35$ times that of the magnetic pressure associated with a $5\mu$G field.  
Equation (14) can be normalized by assuming that the mean radio luminosity
at $r=3.0$ Mpc is $L_{\rm 1.4GHz} = 10^{20.61}$ W~Hz$^{-1}$, according to the best-fit correlation 
(solid line) shown in Figure~\ref{fig:lumvrad}b, and that this value is comparable to the luminosity
of field galaxies.  

For this analysis, the model of radio luminosity expected due to thermal
compression is in general agreement with the observed data.  However, the dashed curve in 
Figure~\ref{fig:lumvrad} should only be considered as representative of the effects that thermal
pressure has on the synchrotron power.  In particular, the normalization of the curve is
subject to a number of factors, including the assumed $B$ field for spiral galaxies and the
assumed $\beta$ model parameters.  The extrapolation of the best-fit curve between 
radio luminosity and radius for the cluster sample out to $3$ Mpc may also be inaccurate as we have
no data that samples the field environment around the clusters.  We also note that the 
intrinsic scatter in the data makes it difficult to determine if the underlying correlation
is truly nonlinear, as expected for thermal pressure effects, or if it is linear, as it
appears in Figure~\ref{fig:lumvrad}b.
Note that the density used in equation (14) is a deprojected quantity.  Therefore, the dashed
curve in Figure~\ref{fig:lumvrad}b increasingly overestimates the radio
luminosity for decreasing radius.  Correcting for this effect will bring the dashed curve downward
toward the solid line for small radii (e.g., within the Core region).  In summary, we find that
thermal compression could explain the relative change in the radio luminosities of
cluster galaxies located within $1.5$~Mpc of cluster centers.

\subsection{Delayed Harassment and Frequency of Radio-Excess Sources}

A recent model of galaxy-galaxy interaction has been proposed by Lake, Katz, \&
Moore (1998).  They carry out numerical simulations of disk galaxies in rich cluster 
environments and show that galaxy ``harassment'' by many galaxy-galaxy interactions can drive up to
90\% of the gas into the inner $500$ pc.  This gas infall can lead to radio enhancement by
(1) fueling AGNs (see \S 6.5.3) and/or quasars in low-luminosity hosts (Lake, Katz, \&
Moore 1998), or (2) nuclear starburst activity (e.g., Fujita 1998).  As discussed previously, 
we can rule out (2) being the {\it sole} mechanism
in producing the correlations discussed in \S 4.3 since it only affects the
star formation rate.  

While harassment is dominant in only the densest cluster environments (e.g., 
cluster cores), the effects of harassment may also influence the Ring population as they
may not fully materialize until some time has passed.  If AGN activation is immediate due to
gravitational, hydrodynamic, or magnetodynamic perturbation, then we should expect
sources with lower $q$ values to have the largest relative velocity.  Figure~\ref{fig:relvel}
suggests the opposite is true, but testing this would require disentangling projection
effects (see \S 6.5.1).

We note that the Core crossing time is $\sim 1.7$ Gyr for a galaxy with a relative cluster 
velocity of $\sim 600$ km~sec$^{-1}$ (similar to the velocity dispersion), much longer 
than the disk crossing time of about $100$ Myr for a starburst.  Consequently, simple 
delayed harassment is an unlikely cause for the observed correlation between $q$ and $r$, as the 
induced activity should dissipate by the time galaxies enter the Ring region.  A more 
complicated model of harassment (e.g., such as one that includes the presence of cluster
substructure) may be needed to explain both the correlation between $q$ and $r$ and the
wide-spread behavior of the mid-IR sources.

The radio-excess fraction among cluster galaxies is $11$ times higher than that among IR-selected
field galaxies.  The sources in the spectroscopically confirmed AGN sample of Miller \& 
Owen (2001a) show a correlation between $q$ and $r$ similar to that of the star forming galaxies,
suggesting that some of the radio emission in AGN sources is due to normal star formation
in the disk of the galaxy (Roy et al. 1998; Baum et al. 1993).
It is therefore plausible that the enhancement in radio emission due to thermal pressure 
is sufficient to increase the radio luminosity of AGN hosts such that 
they fall above the threshold to be considered a radio-excess object.  In particular,
the effects of thermal pressure result in a mean radio luminosity that is 4 times greater
in the Core than in the Ring region, implying a reduction in $q$ value of $\sim 0.6$ dex.  
The $6$ radio-excess sources with $q \gtrsim 1$  may not have been observed as radio-excess 
sources if they were in the field, implying an observed radio-excess fraction of $3/114 \sim 2.6\%$
(excluding cD galaxies), which is still larger than, but consistent with, the radio-excess fraction 
found by YRC ($\sim 1.3\%$).  Note that despite the possible effects of radio enhancement via 
thermal compression in our sample, the large majority of the observed radio-excess sources 
($70\%$) are AGNs.  

\section{Conclusions} 

We have studied the effects of the cluster environment on the star formation and AGN activity
in cluster galaxies.  To accomplish this, radio and FIR data were compiled for a luminosity-limited
sample of 114 cluster galaxies in the most massive, nearby ($<100$ Mpc) X-ray clusters.  The
selection based on X-ray luminosity ensures that we are including only the most massive and
dynamically-related clusters, and is a direct probe of the state of the intracluster gas.  
The seven clusters studied are A262, AWM7, A426, A1060, A1367, Virgo, and Coma.  The main 
conclusions from this study are as follows:

1.  The radio-FIR correlation for the luminosity-limited sample shows almost three times the scatter 
in the correlation for field galaxies.  An examination of the logarithmic FIR-radio flux density
ratio ($q$) reveals that this scatter is mainly due to radio-excess ($q<1.64$) sources, $\sim 70\%$ of
which show evidence of AGN activity.  Including these extreme sources, we find $\langle q\rangle = 2.07$.
The larger cluster sample of Miller \& Owen (2001a) has $\langle q\rangle = 1.95$, lower than our value, and is 
likely due to their selection based strictly on radio emission.
Similar field samples show $\langle q\rangle = 2.34$ (YRC). 

2.  The radio-excess fraction in the Core region of the composite cluster ($R<0.5$ Mpc; $21\%$) 
is larger than that found in the Ring region ($0.5<R<1.5$ Mpc; $7\%$), which is in turn larger than that found in
the field ($1\%$), indicating a correlation with cluster position.  We find a positive correlation
between $q$ and $r$ (clustercentric distance), and this is a result of radio enhancement toward the center of 
clusters.  The Miller \& Owen (2001a) sample also indicates enhanced radio emission by a factor of $2-3$ among 
core galaxies.

3.  We have computed the IR and radio cluster LFs.  The excess of cluster sources above the field radio LF 
at high luminosity, as well as the
excesses observed for the YRC sample and UGC sample all suggest that radio-luminous AGNs contribute more
to the radio luminosity density in the cluster environment than they do in the field.  

4.  The radio and morphological distributions indicate that radio-luminous and early-type galaxies,
respectively, are clustered in the Core region.  We find no evidence that the increase in density is due to a change
in the underlying density-morphology relation due to our radio and FIR selection criteria.  Rather, we are tracing
only a subset of optically-identified early-type galaxies.  Consequently, the increase in early-type density
in the Core region reflects the existing density-morphology relation for all cluster galaxies.

5.  As more massive galaxies may be more efficient radio and IR emitters, we have investigated the possibility
of mass segregation in our sample by compiling $K$-band magnitudes for all candidate sources.  Based on
the analysis of the time-averaged evolution of the composite cluster, we find that massive galaxies are not
preferentially clustered in the Core region.  It is therefore likely that the radio-luminous
elliptical galaxies were formed in the Core region.

6.  An examination of the IRAS colors of the cluster population indicates that radio-excess sources have
warmer mid-IR colors than the normal cluster and field populations, consistent with a model in which dust
is heated in the circumnuclear region by an AGN.  We have independently analyzed the Miller \& Owen (2001a) 
radio-selected sample and find that the frequency of AGNs among their radio-excess sources is $\sim 80\%$. 

7.  ICM electron-dust collisions do not play a significant role in heating the dust in most
cluster galaxies.  The flat FIR luminosity-radius 
relation suggests that environmental effects within the virial radii of clusters do not have a 
significant impact on the FIR emission, despite the factor of $10$ decrease in density from the 
center of the cluster to $1$ virial radius ($\sim 1.5$~Mpc).  

8.  Among the mechanisms resulting in increased radio emission, we cannot rule out a contribution from 
ram pressure enhancement of synchrotron emission.  We find that thermal compression
of a galaxy's ISM by the intracluster medium could also explain the relative enhancement in the radio 
luminosity of centrally-located galaxies compared with those at $1.5$~Mpc.
Thermal compression may increase the synchrotron emission from the disk of a
galaxy and push radio-luminous AGNs into the radio-excess regime.  We argue based on a timescale 
argument that simple delayed harassment is an unlikely cause for the observed correlations, and
a more complicated model including cluster substructure may be needed.

\acknowledgements

This work has greatly benefited from useful comments from, and discussions with, our colleagues 
including Andrew Blain, Chris Conselice, Nick Scoville, Martin Weinberg, and 
the anonymous referee.  We thank Micol Christopher for a thorough proofreading of an
earlier version of this paper.
This work is supported in part by the National Radio Astronomy Observatory (NRAO), 
a facility of the National Science Foundation (NSF) operated under cooperative agreement by 
Associated Universities, Inc.  This publication makes use of data products from the Two 
Micron All Sky Survey (2MASS), which is a joint project of the University of Massachusetts and the 
Infrared Processing and Analysis Center (IPAC), California Institute of Technology, funded by the 
National Aeronautics and Space Administration (NASA) and NSF.
We made use of the NASA/IPAC Extragalactic Database (NED), which is operated by 
the Jet Propulsion Laboratory, California Institute of Technology, under contract 
with NASA, and NASA's Astrophysics Data System Bibliographic Services.  
NAR is supported in part by the NSF Graduate Research Fellowship.

\end{document}